\documentclass[journal = nalefd, layout = traditional]{achemso} 

\setkeys{acs}{usetitle=true}
\setkeys{acs}{maxauthors=20}

\usepackage{graphicx}
\usepackage{hyperref}
\usepackage{soul}
\usepackage{amsmath}
\usepackage[usenames,dvipsnames]{color}
\usepackage{amssymb}
\usepackage{xcolor}

\newcommand{\gap} {\epsilon_L}

\title{Confinement-engineered superconductor to correlated-insulator transition in a van der Waals monolayer}

\author{Somesh Chandra Ganguli}
\email{somesh.ganguli@aalto.fi}
\affiliation{Department of Applied Physics, Aalto University, FI-00076 Aalto, Finland}

\author{Viliam Va\v{n}o}
\affiliation{Department of Applied Physics, Aalto University, FI-00076 Aalto, Finland}

\author{Shawulienu Kezilebieke}
\affiliation{Department of Applied Physics, Aalto University, FI-00076 Aalto, Finland}

\author{Jose L. Lado}
\email{jose.lado@aalto.fi}
\affiliation{Department of Applied Physics, Aalto University, FI-00076 Aalto, Finland}

\author{Peter Liljeroth}
\email{peter.liljeroth@aalto.fi}
\affiliation{Department of Applied Physics, Aalto University, FI-00076 Aalto, Finland}

\begin{document}

\begin{abstract}
Transition metal dichalcogenides (TMDC) are a rich family of two-dimensional materials displaying a multitude of different quantum ground states. In particular, d$^3$ TMDCs are paradigmatic materials hosting a variety of symmetry broken states, including charge density waves, superconductivity, and magnetism. Among this family, NbSe$_2$ is one of the best-studied superconducting materials down to the monolayer limit. Despite its superconducting nature, a variety of results point towards strong electronic repulsions in NbSe$_2$. Here, we control the strength of the interactions experimentally via quantum confinement and use low-temperature scanning tunneling microscopy (STM) and spectroscopy (STS) to demonstrate that NbSe$_2$ is in close proximity to a correlated insulating state. This reveals the coexistence of competing interactions in NbSe$_2$, creating a transition from a superconducting to an insulating quantum correlated state by confinement-controlled interactions. Our results demonstrate the dramatic role of interactions in NbSe$_2$, establishing NbSe$_2$ as a correlated superconductor with competing interactions.
\end{abstract}

\maketitle

\vspace{21pt}
Niobium dichalcogenides, and in particular NbSe$_2$ is well-known to be a paradigmatic superconducting two-dimensional material and it realizes Ising superconductivity at the monolayer (ML) limit \cite{Ugeda2015,Xi2015,Zhao2019}. Due to its superconducting nature, NbSe$_2$ has been considered to be a metal where Coulomb repulsions play a marginal role and the superconducting state arises from conventional electron-phonon coupling \cite{PhysRevLett.92.086401}. Indeed, the emergence of charge density wave states is usually attributed to soft-phonon modes \cite{Soumyanarayanan2013,Lian2018,PhysRevLett.125.106101,Xi2015Nano,PhysRevLett.107.107403,PhysRevLett.122.016403}, so that symmetry broken states are not related with strong Coulomb interactions.

Despite the apparent marginal role of the Coulomb repulsion in NbSe$_2$, related compounds in the dichalcogenide family show strong correlations \cite{OHara2018,Manzeli2017,Wu2018,PhysRevLett.121.196402,PhysRevX.7.041054}. In particular, VSe$_2$ is known to be a strongly correlated material \cite{Duvjir2018}, with competing correlated states including a potential magnetic Mott insulating state \cite{Bonilla2018,Wong2019,Kezilebieke2020,PhysRevLett.121.196402}. The chemical similarity between NbSe$_2$ and VSe$_2$, contrasted with their dramatically different electronic properties, motivates the question of whether NbSe$_2$ exhibits a strongly correlated superconducting state, in contrast with the originally assumed weakly interacting scenario \cite{Zhou2012,PhysRevB.97.081101,divilov2020interplay,PhysRevX.10.041003}. In that regard, theoretical calculations have shown that NbSe$_2$ is close to a Mott insulating transition to a ferromagnetic state \cite{Zhou2012,PhysRevB.97.081101,divilov2020interplay,PhysRevX.10.041003}. These results suggest that competing interactions coexist in NbSe$_2$ system, and in particular suggest the possibility of the superconducting state coexisting with strong Coulomb interactions.

\begin{figure}[t!]
\center
\includegraphics[width=.8\columnwidth]{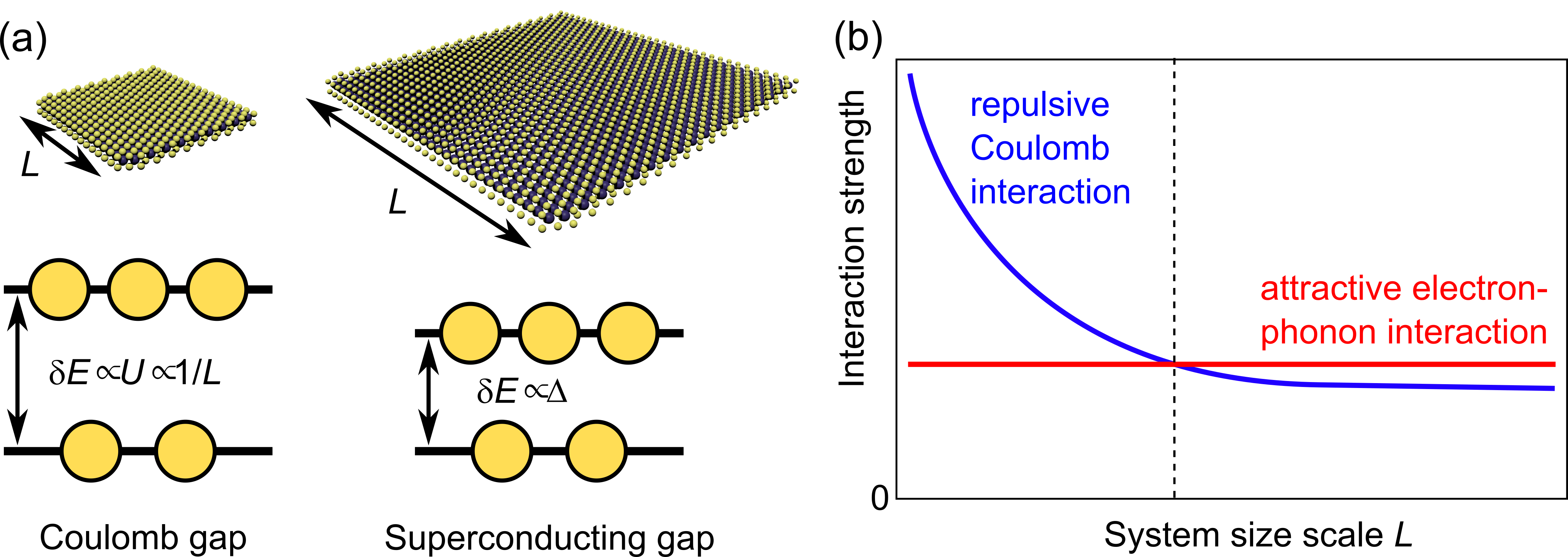}
\caption{(a) Sketches of small and large NbSe$_2$ islands with the associated Coulomb or superconducting gaps. (b) Schematic dependence of the attractive and repulsive interactions on the system size.}
\label{fig:fig1}
\end{figure}

In this manuscript, we experimentally demonstrate that ML NbSe$_2$ is in proximity to a correlated insulating state, by controlling the strength of the electronic interactions by quantum confinement effects. In particular, we show that for ML NbSe$_2$ islands of size several times the coherence length, repulsive electronic interactions create a phase transition from a superconducting to a correlated insulating state. This behavior is rationalized from a competing interaction scenario (Fig.~\ref{fig:fig1}(a)), in which attractive electron-phonon interactions compete with strongly repulsive Coulomb interactions. The electron-phonon interactions that give rise to a superconducting ground state do not depend on the system size, and will dominate if the system size is increased sufficiently (Fig.~\ref{fig:fig1}(b)). On the other hand, the repulsive Coulomb interactions are strongly dependent on the system size ($U\propto 1/L$ \cite{Kouwenhoven2001,Guinea2018,Nuckolls2020}) and will drive the system into Coulomb-gapped, correlated state as the system size is decreased. This picture is complementary to the classical interpretation in terms of Coulomb blockade, and completely analogous to the approach taken in, e.g., interpreting the correlated insulating states in twisted bilayer graphene in terms of local repulsion \cite{Nuckolls2020}. We test this behaviour experimentally by tuning the size of NbSe$_2$ islands and using low-temperature scanning tunneling microscopy (STM) and spectroscopy (STS) to measure the type and magnitude of the resulting energy gap. Our results provide a quantitative experimental bound on the strength of repulsive interactions of NbSe$_2$, highlighting a non-trivial impact of correlations in superconducting dichalcogenides.

\begin{figure*}[t!]

\includegraphics[width=0.99\linewidth]{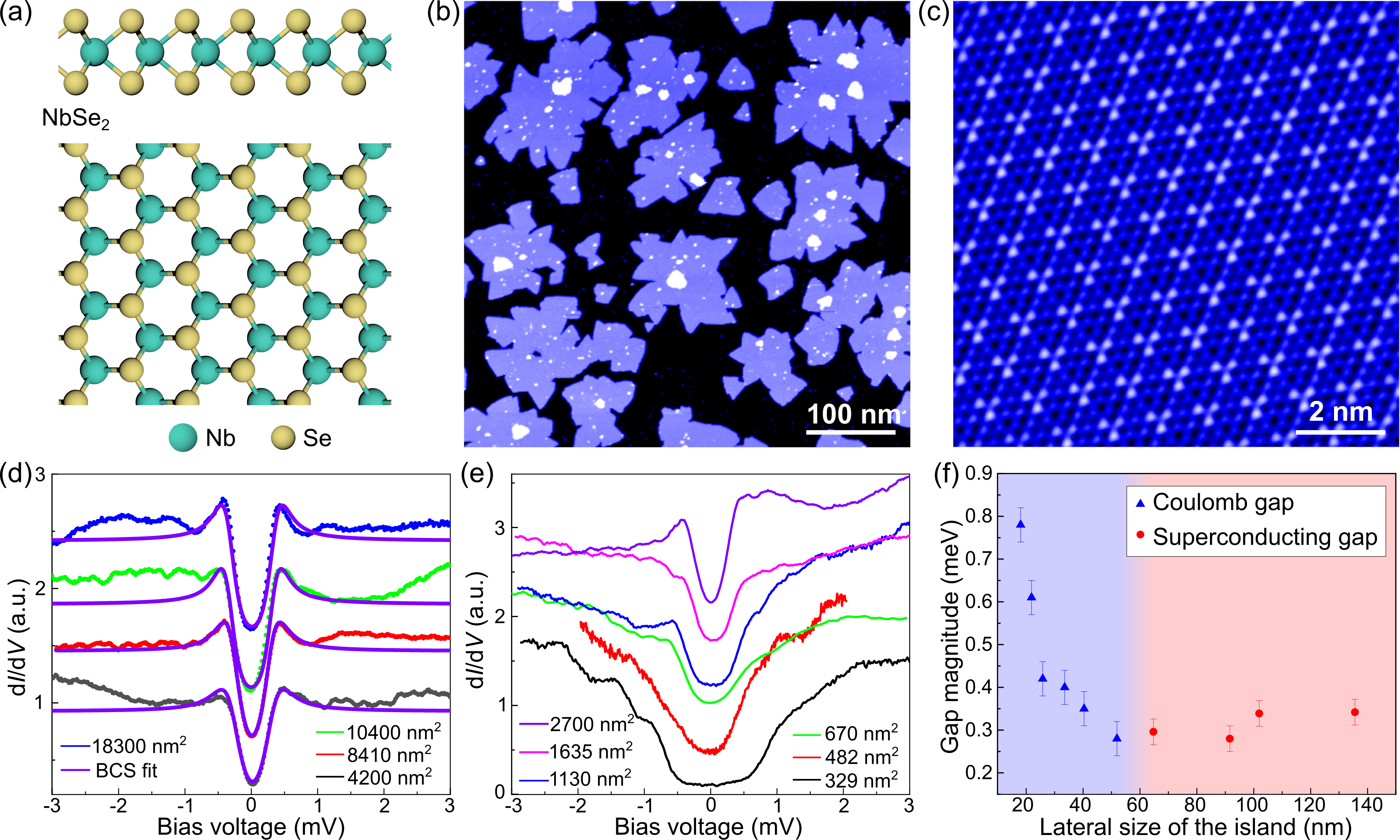}
\caption{(a) Side and top view schematics of monolayer NbSe$_2$. (b) Large-scale STM image of monolayer NbSe$_2$ on HOPG showing a large variation of island sizes. (c) Atomic resolution image of monolayer NbSe$_2$ showing $3\times3$ charge density wave modulation. (d, e) Variation of the superconducting (d) and Coulomb (e) gap with island size. Spectra have been normalized and offset vertically. Superconducting gaps in panel (d) have been fit with the Dynes equation (solid purple lines). (f) Evolution of the gap magnitude extracted from the tunneling spectra as a function of the island size showing a transition from Coulomb gap-like to superconducting spectra as the size is increased. The shape of the measured spectrum is indicated by the different symbols: blue triangles and red circles for the spectra exhibiting Coulomb- and SC-type gaps, respectively.}
\label{fig:fig2}
\end{figure*}


\textbf{Experimental superconducting-correlated transition.}
We grow NbSe$_2$ (Fig.~\ref{fig:fig2}(a)) on a highly-oriented pyrolytic graphite (HOPG) substrate with a sub-monolayer coverage. By adjusting the growth conditions (see Supporting Information (SI) for details), we achieve a sample with a wide variety of island sizes and their relative separations. This creates an ideal platform to study the effects of quantum-confinement enhanced correlations. The island sizes vary between a few hundreds of nm$^2$ to several tens of thousands of nm$^2$ (lateral sizes few tens of nm to several hundreds of nm, see SI for island size determination). Fig.~\ref{fig:fig2}(b) shows an STM image of a representative area ($500\times500$ nm$^2$), where this size variation of individual ML islands is apparent. Each individual island has atomically sharp edges and show the well-known $3\times3$ charge density wave (CDW) modulation similar to extended ML NbSe$_2$ (Fig.~\ref{fig:fig2}(c)). While the data shown in Fig.~\ref{fig:fig2}(c) was acquired on a NbSe$_2$ island with a lateral size of $\sim92$ nm (area 8400 nm$^2$), the CDW modulation persists down to islands sizes of $<500$ nm$^2$ (see SI). We characterize the electronic properties of each individual island by carrying out spatially resolved tunneling conductance (d$I$/d$V$) measurement (see Methods section for details). Typical examples of the d$I$/d$V$ spectra are shown in Figs.~\ref{fig:fig2}(d) and \ref{fig:fig2}(e). The spectra can be divided into two groups based on qualitative differences. Islands with sizes 4200 nm$^2$ and above show density of states consistent with BCS-like behaviour with particle-hole symmetric coherence peaks (Fig.~\ref{fig:fig2}(d)), which indicate a presence of phase-coherent Cooper pairs. On the other hand, islands with sizes 2700 nm$^2$ and below have distinctive particle-hole asymmetric density of states (Figure~\ref{fig:fig2}(e)) with no coherence peaks. This transition occurs at a size range several times larger than the coherence length of NbSe$_2$ ($\sim 7$ nm, see below).

Such asymmetric differential conductance is typical of inelastic steps associated to correlated Coulomb excitations  \cite{Heinrich2004,Khajetoorians2010,Bryant2015}. Furthermore, the magnitude of the energy gap in these islands monotonically increases with decreasing island size (Fig.~\ref{fig:fig2}(f), the details of extracting the energy gap are given in the SI) \cite{Kouwenhoven2001,ihn2010semiconductor}. This behaviour is consistent with the presence of a Coulomb gap in small islands, where the repulsive Coulomb interaction dominates over phonon-mediated attractive interactions. On the other hand, the BCS-shaped superconducting gaps in the islands in Fig.~\ref{fig:fig2}(d) are  independent of the island size (Fig.~\ref{fig:fig2}(f)) as the electron-phonon coupling strength does not depend on the system size. The Coulomb gap and superconducting gaps can also be distinguished by their respective magnetic field dependent behaviour (see SI).

Disorder driven superconductor-insulator transition (SIT) cannot justify our data, since we observe that the SC gap remains practically constant with reducing island size. The data also suggests that disorder does not have a sizable impact besides a small renormalization of the density of states as given by the Altshuler-Aronov effect\cite{altshuler1985electron,PhysRevB.100.045413}. The main source of disorder appears to be on the edges of the islands which show slightly different SC energy gap compared to the center of the island (see SI).

\textbf{Theoretical model for competing interactions.}
However, the previous phenomenology can be rationalized with a many-body low energy model. 
Many-body interactions are well known to lead to Coulomb blockade effects in conventional superconductors,
promoting intriguing phenomena arising from the interplay of pairing correlations and finite size effects\cite{Bose2014,Yuan2020}. However, these phenomena has remained unexploited to probe many-body effects in correlated two-dimensional superconductors.
Since the full quantum many-body system for a nm-sized island cannot be exactly solved, we will focus on the instability of the lowest energy $2n$ single-particle eigenstates of the NbSe$_2$ island $\Psi_{i,s}$, with $i=1,..,n$ the state number and $s={\uparrow,\downarrow}$ the spin quantum number.
These states closest to the Fermi energy will be the ones most impacted
by interactions, and therefore the fundamental physics
of the system can be captured
by projecting electronic interactions in this manifold.
For the sake of concreteness, we take interactions
$SU(2)$ symmetric
and constant on the Fermi surface manifold.
In particular, we take projected electronic interactions
partitioned into intra-orbital repulsive ones $U$ (of Coulomb origin)
and inter-orbital
attractive ones $V$ (of electron-phonon origin).
Furthermore, due to the existence of nearby large superconducting islands,
the low energy states will feel a superconducting
proximity effect with a value depending on the distance to the closest big superconducting island. We parametrize this effect with $\bar \Delta$.
The half filling of the low energy manifold is enforced
by $\mu$, and computed self-consistently for each $U$ and $V$.
The low energy many-body Hamiltonian takes the form.
\begin{equation}
\begin{split}
    \mathcal{H} = &
    \sum_{i} U 
    \Psi^\dagger_{i,\uparrow} \Psi_{i,\uparrow}
    \Psi^\dagger_{i,\downarrow} \Psi_{i,\downarrow} \\
    &
    -\sum_{i,j>i,s,s'} V 
    \Psi^\dagger_{i,s} \Psi_{i,s}
    \Psi^\dagger_{j,s'} \Psi_{j,s'} \\
    &
    + \mu \sum_{i,s} \Psi^\dagger_{i,s} \Psi_{i,s}
    +  \bar \Delta \sum_i \Psi^\dagger_{i,\uparrow}\Psi^\dagger_{i,\downarrow}
    + \text{H.c.}
    \end{split}
    \label{eq:h}
\end{equation}

The projected
electron-phonon interaction $V$ is taken to be independent on the system size,
whereas the projected Coulomb repulsive interaction $U$ will get enhanced as the system size $L$
becomes smaller as $U= U_0 + \frac{c_0}{L}$ due to the long-range
tail of Coulomb interactions.
The effective model is solved using exact diagonalization, projecting the electronic
repulsion onto the lowest energy states, and solving the
projected Hamiltonian exactly. This is, of course, an approximate procedure when a finite
number of states is considered, and we verified
that our results are not qualitatively modified
when including a higher number of orbitals. 
For such a many-body Hamiltonian the single-electron density of states
can be computed as $A(\omega) = \sum_{i,s} \langle \Omega | \Psi_{i,s}  \delta (\omega - \mathcal{H} + E_0)
\Psi^\dagger_{i,s} |\Omega \rangle$, where $E_0$ is the many-body energy
and $|\Omega\rangle$ the many-body ground state.

\begin{figure}[t!]
\center
\includegraphics[width=0.8\columnwidth]{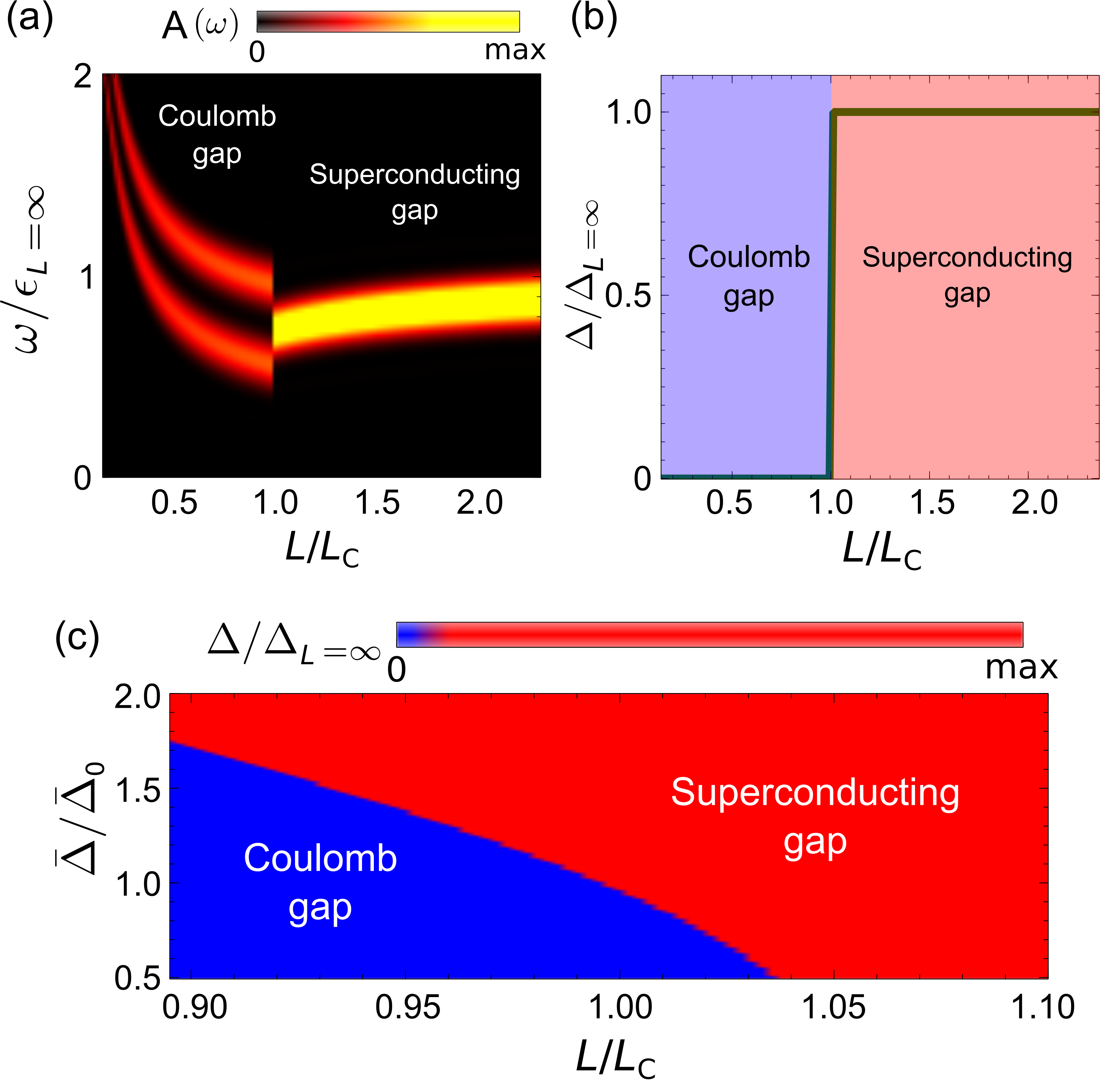}
\caption{(a,b,c) Electron spectral function as a function
of the system size (a), and induced superconductivity as a function of the system size (b). 
It is shown that a transition between the superconducting and the correlated state takes place without gap closing.
For correlated islands close to the phase transition,
increasing the superconducting proximity effect $\bar \Delta$
can push the system to the superconducting
region as shown in panel (c.
We took $2n=8$ for A-C, $U_0=2$ V and $L_C$
is the critical length for $\bar \Delta_0 =0.4$ V.}
\label{fig:fig3}
\end{figure}

We show in Fig. \ref{fig:fig3}(a) the single-electron
spectral function $A(\omega)$
as a function of the system size $L$,
where the transition between a Coulomb dominated gap $\gap$ to a superconducting dominated one
can be seen.
For large system size $L\rightarrow \infty$, 
the system shows a superconducting gap stemming from the attractive interactions and pinned by the
superconducting proximity $\bar \Delta$. 
It is worth noting that, as the system is finite, observing a
sharp phase transition from zero to finite
superconducting order
requires a finite value of the proximity effect $\bar \Delta$.
Once the system size goes below a critical value $L_C$, the
nature of the excitation gap $\gap$ changes, yet without a gap closing. 
The different nature of
the two gaps above and below the transition point $L=L_{C}$ can be verified by computing the superconducting
expectation value
$\Delta = \sum_i \langle \Psi_{i,\uparrow} \Psi_{i,\downarrow} \rangle$, showing that associated with the
discontinuous jump as the size becomes smaller,
the superconducting order parameter suddenly disappears (Fig.~\ref{fig:fig3}(b)). 
We note that for small islands the observed spectra featuring a continuum of states above the gap are fundamentally different from the ones expected for systems with confined energy levels. The transition between
the correlated gap for small islands and superconducting one for large islands is found to be of first order, with a discontinuity on the gap. This is consistent with our experimental data and therefore strongly supports the competing interaction scenario.

Due to the proximity of NbSe$_2$ to the phase transition point, it is expected
that an external perturbation can cause a critical system to drift to different
regions of the phase diagram. In particular, increasing a superconducting
proximity effect $\bar \Delta$ would push the system toward the superconducting gapped
region. This can be verified
as shown in Fig. \ref{fig:fig3}(c) 
where it can be seen that ramping up the superconducting
proximity pushes the system that
originally has a correlated gap
towards a superconducting gap. While this is shown for reduced range in Fig.~\ref{fig:fig3}(c), the
very same mechanism applies in a broader range of $L$ and $\Delta$. We have verified that the same behavior
remains qualitatively unchanged upon increasing the number of orbitals considered in the many-body Hamiltonian (shown in the SI).

It is well-known that 2H-NbSe$_2$ exhibits charge-density wave order at low temperatures and the presence of Ising-type spin-orbit coupling might also have an effect on the observed behaviour \cite{Xi2015,PhysRevLett.92.086401,PhysRevLett.125.106101,Xi2015Nano,delaBarrera2018}. However, by using a more detailed model incorporating these two effects (see SI), we can demonstrate that the observed phenomenology is a genuine Coulomb effect. Ising spin-orbit coupling leads to momentum dependent spin splitting in the Brillouin zone. As this perturbation respects time-reversal symmetry, it does not have a detrimental impact on spin-singlet superconducting state. Ising SOC will also not impact the Coulomb electronic interactions, as the momentum-dependent exchange splitting does not change
the underlying atomic nature of the orbitals, which is the one that ultimately determines the strength of the local and non-local interactions.

The NbSe$_2$ charge density wave gives rise to band folding and splitting, yet maintaining the system metallic. Even though the
low energy states now become modulated in space following the CDW profile, the repulsive Coulomb interactions are not qualitatively affected. This is rationalized from the fact that electronic repulsion is an atomic property associated to the Wannier states, and thus again independent on larger scale structural reconstructions. The CDW also does not substantially impact the mechanism for superconductivity, since such reconstruction does not break time-reversal symmetry, the relevant symmetry that could strongly impact the spin-singlet superconducting state.

\begin{figure*}[t!]
\center
\includegraphics[width=.99\linewidth]{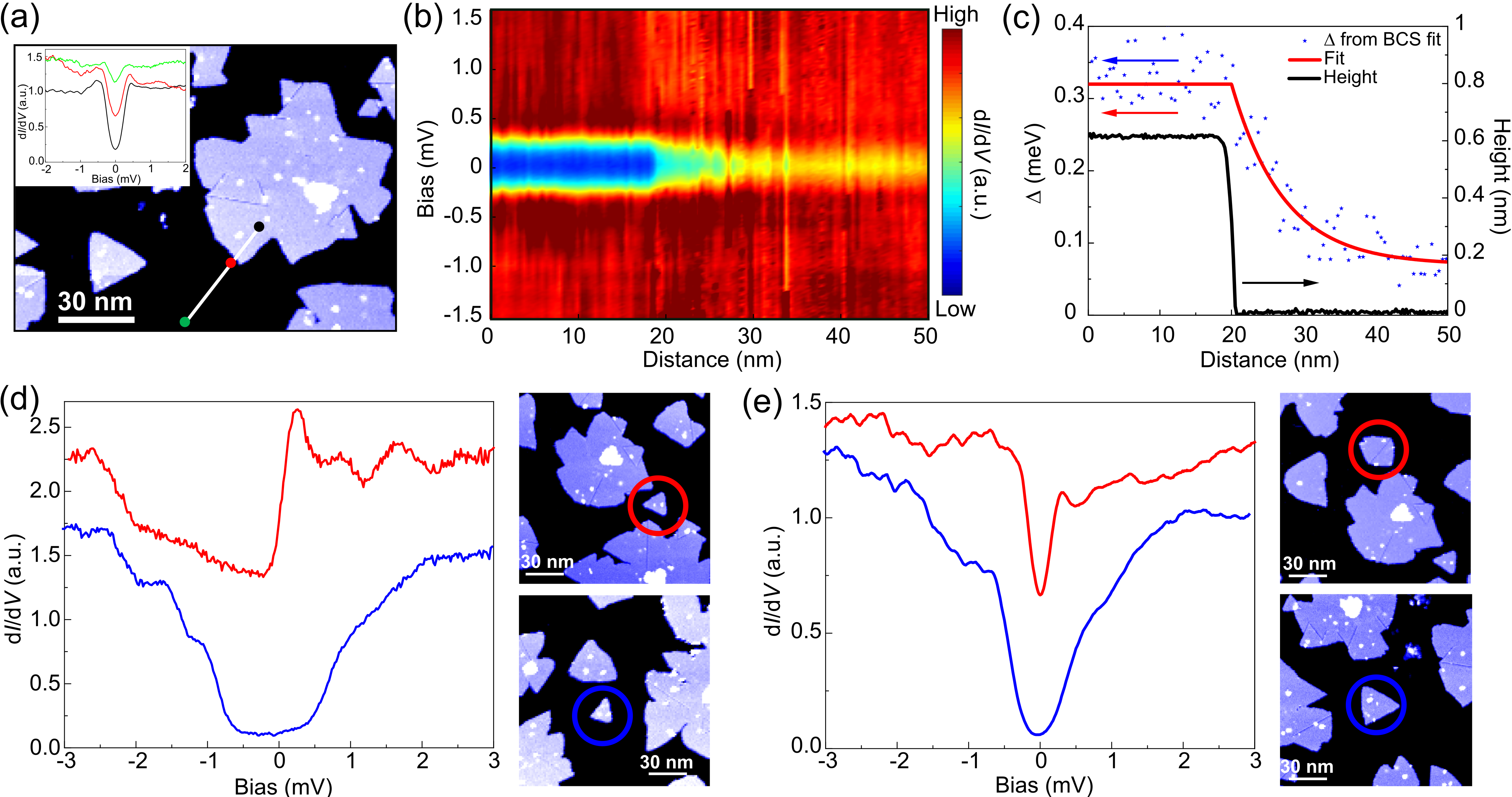}
\caption{(a) STM image showing NbSe$_2$ monolayer island. d$I$/d$V$ spectra measured in  black, green and red points are shown in the inset with corresponding colors. (b) d$I$/d$V$ spectra measured along the white line in panel (a) presented as a color scale plot. Black (green) point in (a) is the left (right) edge of (b). (c) Fitted SC gap, its exponential fit along with the height profile measured along the white line in panel (a). (d,e) Proximity induced superconductivity in Coulomb gapped islands. d$I$/d$V$ spectra and topographic images of (d) an isolated island of size 330 nm$^2$ (blue circle) and an island of size 330 nm$^2$ in proximity with larger SC island (red circle), and (e) an isolated island of size 650 nm$^2$ (blue circle) and an island of size 650 nm$^2$ in proximity with larger SC island (red circle). Scale bars, 30 nm. Spectra in panels (d) and (e) are offset vertically for clarity.}
\label{fig:fig4}
\end{figure*}

\textbf{Proximity induced quantum phase transition.}
Based on the previous results, we check this proximity-induced phase transition experimentally by comparing the spectra of different critical islands with different respective distances to a big superconducting island, probing whether the superconducting proximity effect transforms the correlated gap into a superconducting one. We start by quantifying the proximity effect in the NbSe$_2$/HOPG-system as shown in Figs.~\ref{fig:fig4}(a-c). Measuring d$I$/d$V$ spectra close to a SC NbSe$_2$ shows a proximity-induced gap on HOPG and tracking the spatial evolution allows us to estimate the decay length. Fitting the spatially dependent d$I$/d$V$ in Fig.~\ref{fig:fig4}(b) to Dynes equation, we extract the gap as a function of the distance from the NbSe$_2$ island edge (Fig.~\ref{fig:fig4}(c)). An exponential fitting of Dynes gap with distance yields $\xi \approx 7$ nm (see SI).

We then proceed to show the effect of proximity in the non-superconducting islands showing size-dependent Coulomb gaps. We selected 2 representative island sizes of 330 nm$^2$ and 650 nm$^2$ (Fig.~\ref{fig:fig4}(d,e), additional results on spatially resolved spectroscopy are shown in the SI). Here, the smaller of the islands is well into the Coulomb gapped regime, but the larger one is closer to phase transition determined in Fig.~\ref{fig:fig2}(f). When each of these islands are not in proximity ($\sim7$ nm) to any superconducting islands (Figs.~\ref{fig:fig4}(d,e)), they show particle-hole asymmetric Coulomb gap (Figs.~\ref{fig:fig4}(d,e), blue lines). Island with size 650 nm$^2$ in proximity to a larger superconducting island shows a drastically different conductance with gap value comparable to the BCS gap observed in larger islands (Fig.~\ref{fig:fig4}(e), red line), indicating that the proximity effect is sufficient to push the system into the superconducting phase. Strong particle-hole asymmetric feature indicates significant presence of correlation in this proximity-induced superconducting island. The magnetic field dependent behaviour of proximitized island is also indicative of the presence of superconducting order (see SI). On the other hand, island with size 330 nm$^2$ in proximity to a larger superconducting island shows a complex spectra with no clear gap signature (Fig.~\ref{fig:fig4}(d), red line), indicating that the proximity-induced Josephson coupling is not sufficient to overcome Coulomb repulsion to induce superconducting order in this island. 

\textbf{Conclusions.} We have demonstrated that ML NbSe$_2$ can be pushed to a correlated
regime, driving a quantum phase transition from superconducting to a correlated gap. This transition is
rationalized from the existence of competing interactions, 
in which the coexistence of attractive electron-phonon
interactions, driving superconductivity,
and repulsive Coulomb interactions, driving correlated insulating behavior, 
allows to dramatically
change the nature of the ground state in NbSe$_2$ 
by slightly enhancing the Coulomb interactions. The Coulomb gap observed in our system is inherently different from the single-particle gap observed in small metallic islands \cite{von2001spectroscopy}. The d$I$/d$V$ spectra in the smallest NbSe$_2$ islands (Fig.~\ref{fig:fig2}(e)) show a continuum of states above the gap rather than a discrete set of states \cite{schneider2013coulomb}, indicating many-body nature of the gap in our system. While it is possible to analyze our data using the Coulomb blockade model typically employed for 3D superconductors \cite{lafarge1993measurement,brun2012dynamical,vlaic2017superconducting}, it is worthwhile to note that these systems are weakly interacting being far from any Stoner instability and electron induced symmetry breaking, whereas NbSe$_2$ is in close proximity to correlated state which can be driven by perturbations such as strain \cite{PhysRevX.10.041003}. Also, the SIT mechanism observed here veers away from the traditional disorder-driven scenario. In comparison, similar SIT has been observed by controlling electronic interactions in twisted van der Waals multilayers \cite{Stepanov2020}.

The critical role of Coulomb interactions highlighted
in our results
suggests a potentially crucial impact of electronic correlations
for the emergence of both charge density wave orders and superconductivity
besides the typical electron-phonon driven scenarios. Recent results show the presence of spin-fluctuations in ML NbSe$_2$ \cite{wan2021observation} and nematic superconductivity in few layer NbSe$_2$ \cite{Hamill2021}, which are indicative of its proximity to correlated regime.
We finally showed that for correlated NbSe$_2$ samples close to the phase transition,
superconducting proximity effect strongly impacts the ground state,
pushing the system through the superconductor-correlated phase boundary.
Ultimately, these results suggest that due to the
close to critical behavior of NbSe$_2$,
correlated states could be promoted in NbSe$_2$
by screening \cite{Stepanov2020}, chemical \cite{Zhang2017} or
twist engineering \cite{Shimazaki2020}, putting forward
$d^3$ chalcogenides as paradigmatic
strongly correlated two-dimensional 
materials.

\begin{suppinfo}
Experimental methods, and additional experimental and theoretical results. 
\end{suppinfo}

\begin{acknowledgement}
This research made use of the Aalto Nanomicroscopy Center (Aalto NMC) facilities and was supported by the European Research Council (ERC-2017-AdG no.~788185 ``Artificial Designer Materials'') and Academy of Finland (Academy professor funding nos.~318995 and 320555, Academy postdoctoral researcher no.~309975, Academy research fellow nos.~331342 and 336243). 
We acknowledge the computational resources provided by the Aalto Science-IT project. We thank Xin Huang and H\'ector Gonz\'alez-Herrero for their help with the temperature dependence of the NbSe$_2$ growth.
	
\end{acknowledgement}

\bibliography{biblio}{}

\end{document}


\section*{Methods}

\textbf{MBE growth.}
Sub-monolayer NbSe$_2$ was grown by molecular beam epitaxy (MBE) on highly oriented pyrolytic graphite (HOPG) under ultra-high vacuum conditions (UHV, base pressure $\sim1\times10^{-10}$ mbar). HOPG crystal was cleaved and subsequently out-gassed at $\sim400^\circ$C. High-purity Nb and Se were evaporated from an electron-beam evaporator and a dual-filament low temperature Knudsen cell, respectively. The flux ratio of Nb to Se was controlled to be $\sim1:30$. During the growth the substrate temperature was kept at $\sim330^\circ$C, and after the growth the sample was annealed at the same temperature for 1 hour. The growth speed was determined by checking the coverage of the as-grown samples by scanning tunneling microscopy (STM). For higher temperature growth ($>$ $500^\circ$C), we obtain significant proportion of 1T-NbSe$_2$ islands.

\textbf{STM/STS measurements.}
Subsequent to the growth, the sample was transferred to a low-temperature STM (Unisoku USM-1300) housed in the same UHV system. STM imaging and STS experiments were performed at $T=350$ mK. STM imaging was performed in constant current mode. Differential conductance (d$I$/d$V$) spectra were measured using standard lock-in techniques sweeping the sample bias in an open feedback loop with rms bias modulation of 70 $\mu$V at a frequency of 873.7 Hz.

\begin{figure*}[h!]

\includegraphics[width=0.65\linewidth]{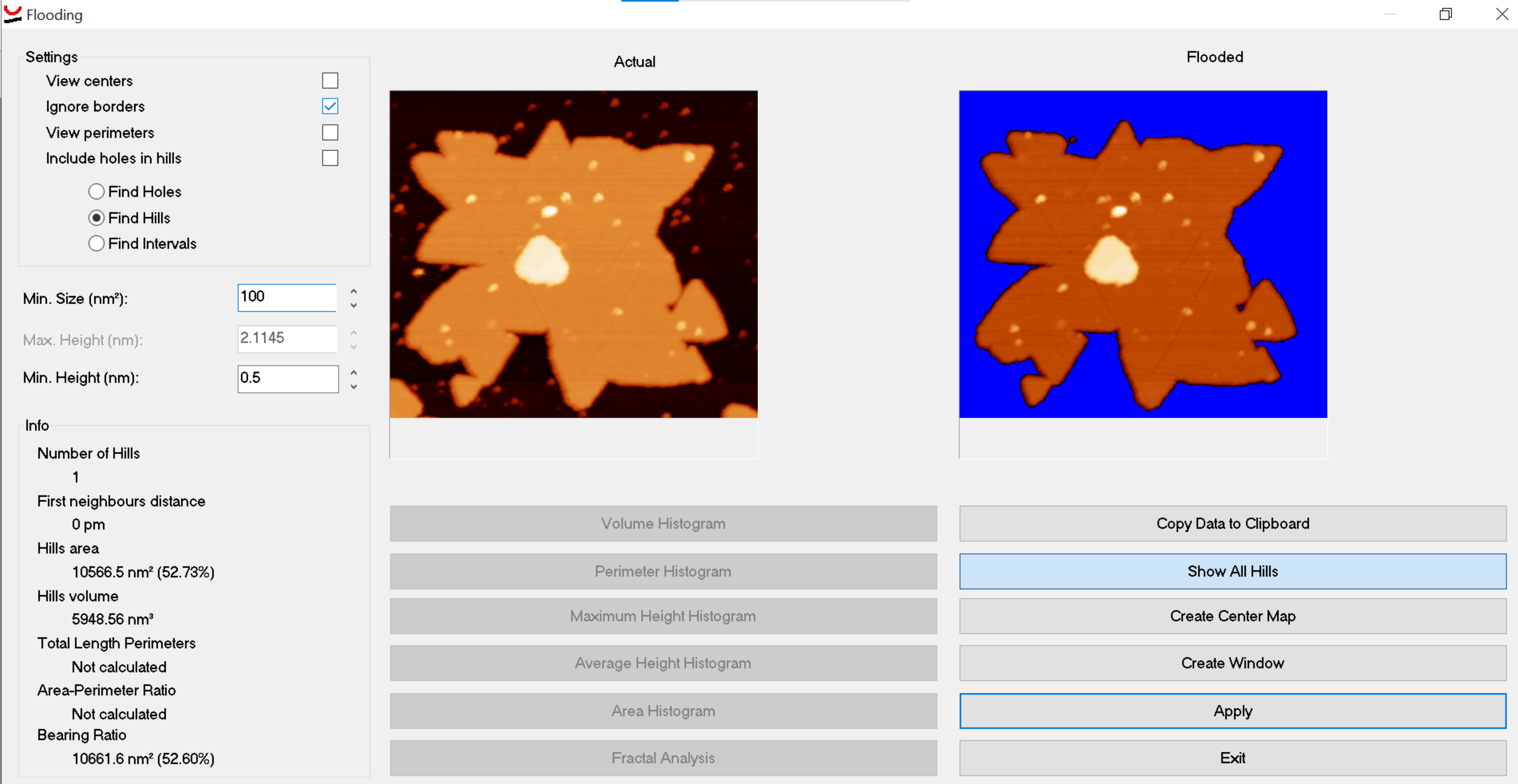}
\caption{Determination of monolayer NbSe$_2$ island size using WSxM software.}
\label{fig:fig1}
\end{figure*}

\textbf{Determination of the island size.}
The size of the islands were determined using WSxM software \cite{horcas2007wsxm}, where the area of the feature size having height $\sim 6$ \AA~  (monolayer height) was used as the area of a monolayer island (Fig.~\ref{fig:fig1}). Lateral size was determined as $\sqrt{area}$.

\textbf{Determination of the SC and Coulomb gap magnitudes.}
For determining the average SC gap, spectra were averaged over a $12.5\times12.5$ nm$^2$ area $10$ nm away from the edge of the island. The data in Fig.~2 was recorded on islands that were non-proximitized (i.e.~had no other islands in their immediate vicinity).

\begin{figure}[t!]
\center
\includegraphics[width=0.75\linewidth]{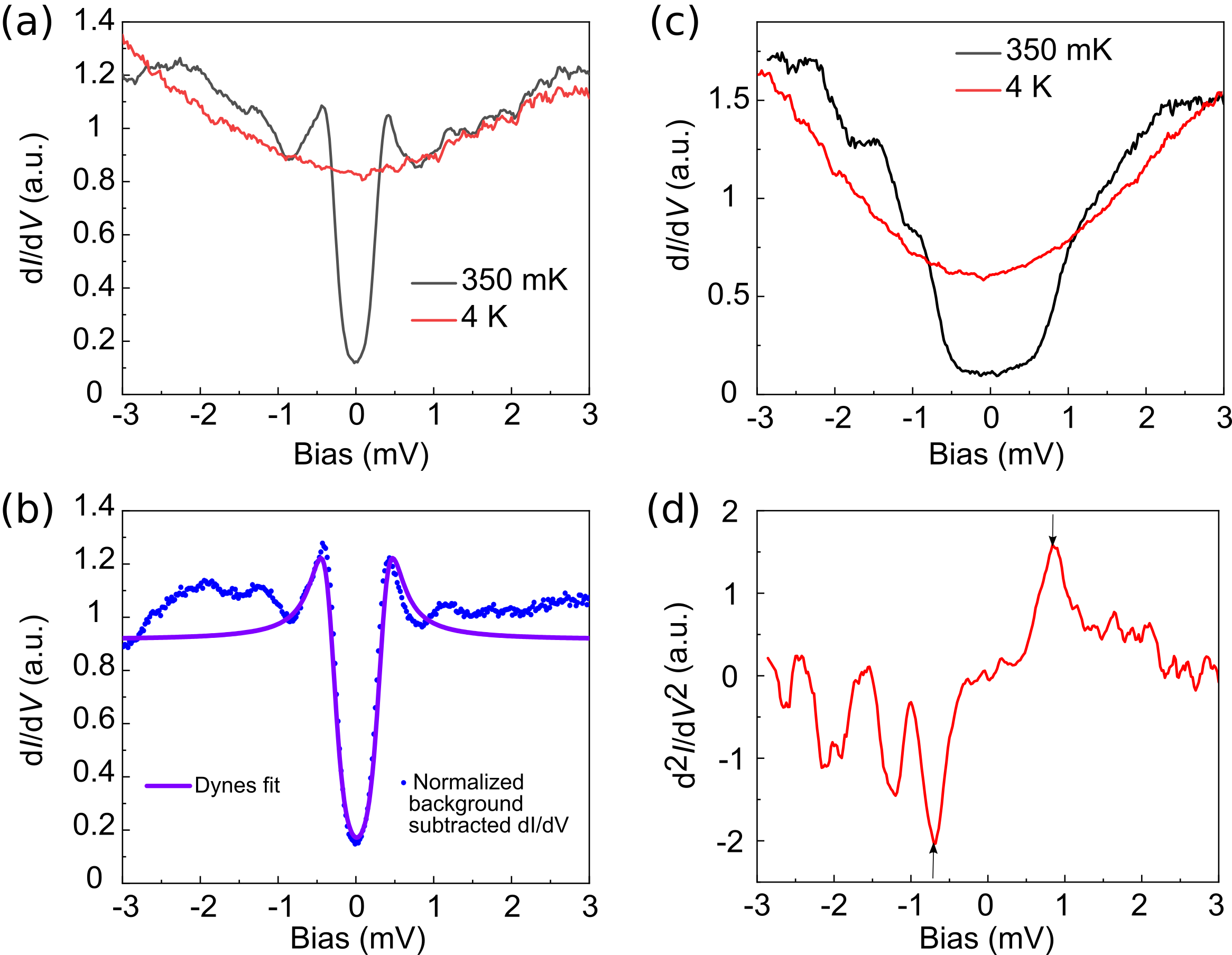}
\caption{Background removal and fit of SC spectra and estimation of Coulomb gap: (a) Normalized 350 mK and 4 K spectra, taken in 18400 nm$^{2}$ island. (b) 350 mK spectrum after removing the 4 K background along with its Dynes fit. (c) Coulomb gap spectra taken in 329 nm$^{2}$ island (spectrum at $T=4$ K shown for reference. (d) Numerical derivative  $\mathrm{d}^{2}I/\mathrm{d}V^{2}$ of the 350 mK spectrum (arrows indicate Coulomb gap energies).}
\label{fig:AAback}
\end{figure}

\begin{figure}[t!]
\center
\includegraphics[width=0.45\linewidth]{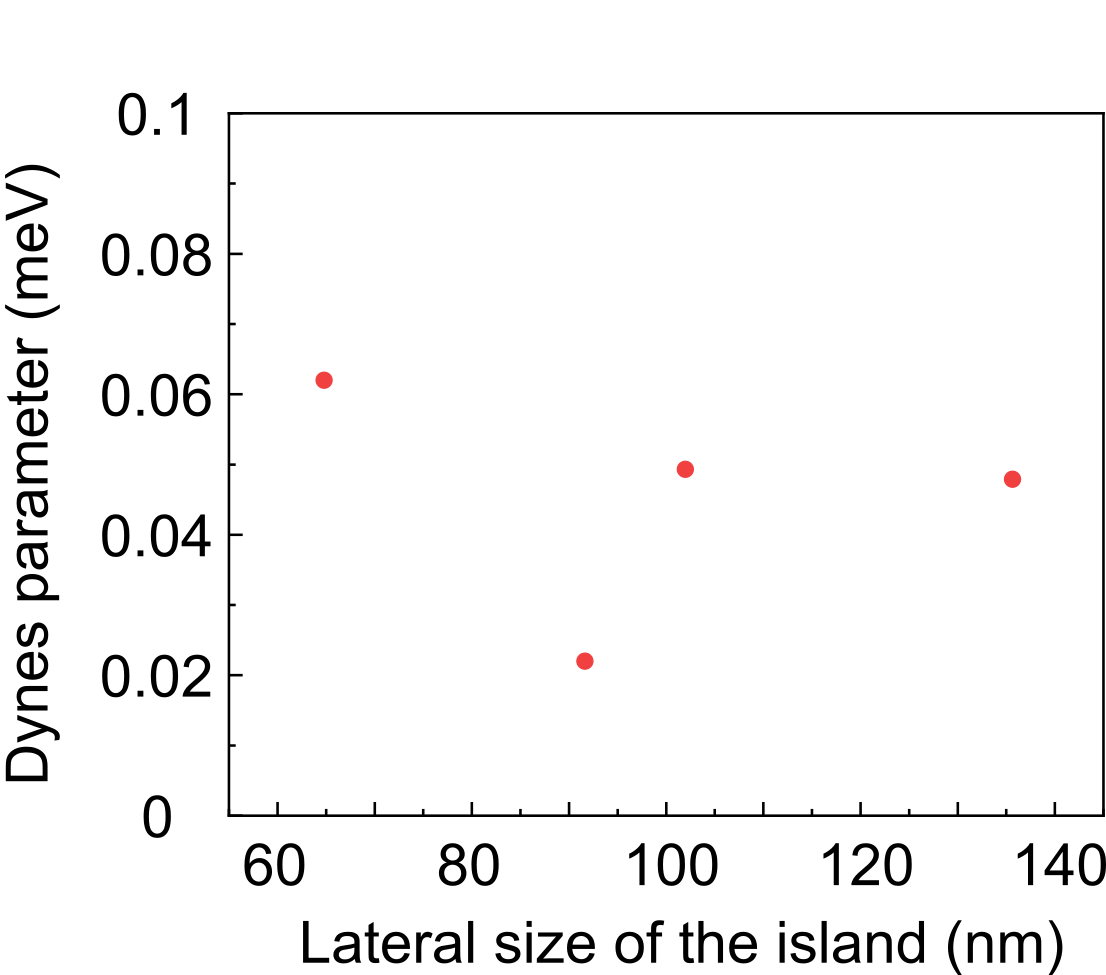}
\caption{Size dependence of Dynes parameter from the fit with respect to the size of the SC islands.}
\label{fig:Dynes}
\end{figure}

At first, the d$I$/d$V$ spectra on the superconducting islands were normalized with respect to the normal state conductance. These normalized d$I$/d$V$ spectra had a temperature independent V-shaped background at above superconducting $T_{c}$ which was also apparent in the lowest temperature (at 350 mK) spectra as shown in Fig.~\ref{fig:AAback}(a). This is the well known Altshuler-Aronov effect due to electronic interactions. So, this background was subtracted by dividing the superconducting spectra with the 4K spectra and the resultant spectra was fitted to the Dynes function $N_{S}(E)=\Re(\frac{\abs{E}+i\Gamma(T)}{\sqrt{(\abs{E}+i\Gamma(T))^{2}-\Delta(T)^{2}}})$ as shown in Fig.~\ref{fig:AAback}(b). Here, $N_{S}(E)$ denotes the normalized background subtracted SC density of states, $\Re$ denotes the real part, $\Gamma(T)$ is the temperature dependent quasiparticle lifetime broadening parameter, also known as the Dynes parameter and $\Delta(T)$ is the temperature dependent SC energy gap. The size dependence of the extracted Dynes parameter is shown in Fig.~\ref{fig:Dynes}.

To extract the Coulomb gap magnitude the raw spectra (Fig.~\ref{fig:AAback}(c)) was numerically differentiated to obtain d$^2I$/d$V^2$, which has a characteristic peak-dip features in positive and negative biases respectively, as shown by arrows in Fig.~\ref{fig:AAback}(d). The average bias voltage location for peak and dip was taken to be the Coulomb gap.

\section{Size dependence of CDW and band structure}
\begin{figure}[h!]
\center
\includegraphics[width=0.8\linewidth]{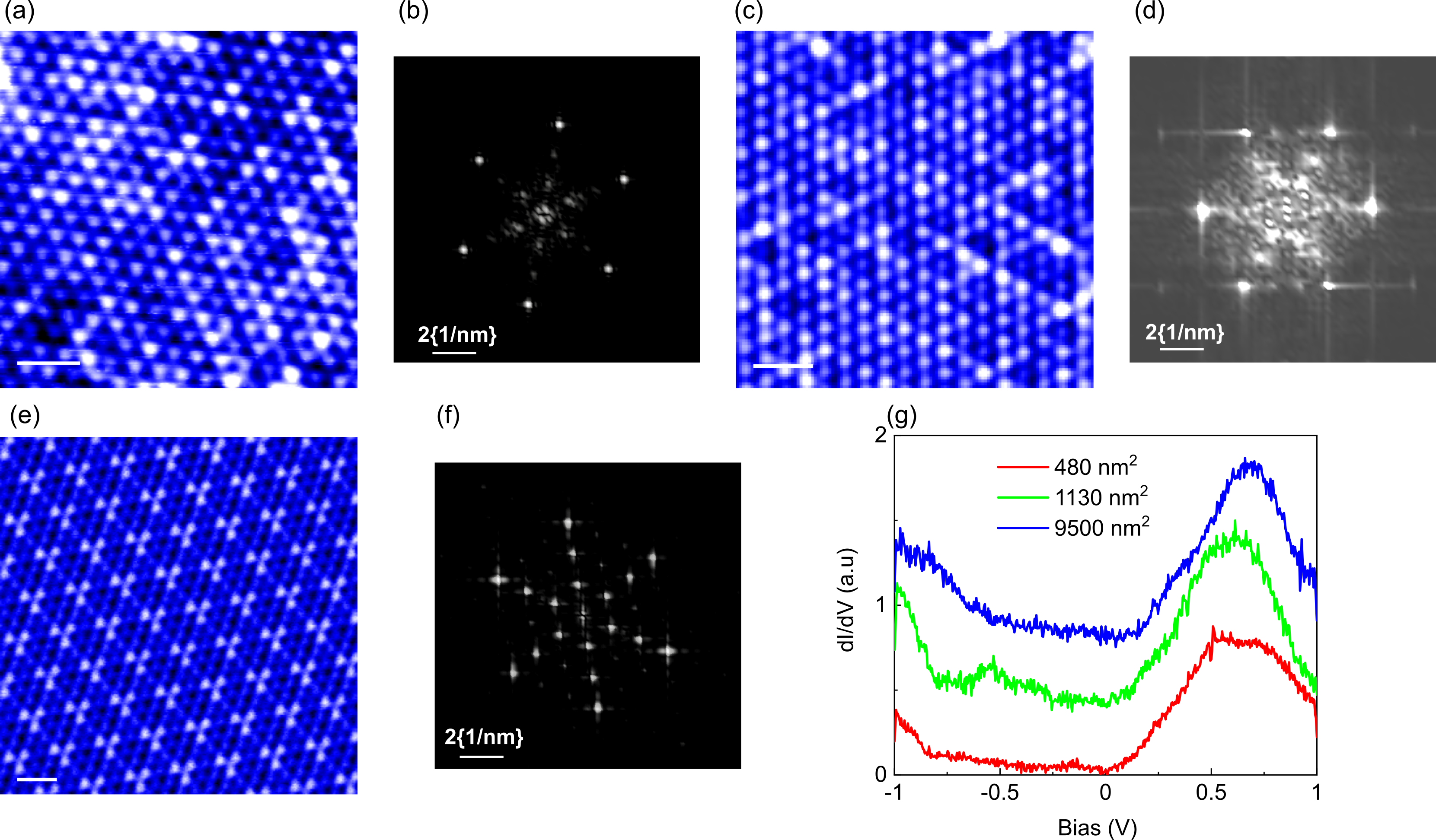}
\caption{Size dependence of CDW and band structure.  Examples of STM images and the corresponding FFTs showing the presence of the CDW state for NbSe$_2$ islands of different sizes: (a),(b) Island size 480 nm$^2$, (c),(d) Island size 1130 nm$^2$, (e),(f) Island size 9500 nm$^2$. Scale bars 1 nm. (g) Large bias range d$I$/d$V$ spectra showing the presence of the Nb d-band at the Fermi level for all island sizes.}
\label{fig:Size dep}
\end{figure}

In Fig.~\ref{fig:Size dep}(a,c,e), we show the presence of 3 $\times$ 3 charge density wave in islands of sizes 480 nm$^2$, 1130 nm$^2$ and 9500 nm$^2$ respectively as indicated by their respective fast Fourier transforms (FFT) in Fig.~\ref{fig:Size dep}(b,d,f). This indicates that the  $3\times3$ CDW modulation present in the extended monolayer NbSe$_2$ survives down to the length scales where correlations are observed. Also, the intensity of the CDW modulation becomes inhomogeneous in the correlated regime. This is a further signature of the strengthening of interactions; perhaps such a change could be associated to an additional charge ordering promoted by interactions. In Fig.~\ref{fig:Size dep}(g), the large bias d$I$/d$V$ remains unchanged with island sizes showing the characteristic Nb d-band feature observed in the extended monolayer. All these results clearly demonstrates that the 2H-polytype of NbSe$_2$ survives down to the length scales where correlations are being observed. 

\pagebreak

\newpage

\section{Magnetic field dependence}

\begin{figure}[h!]
\center
\includegraphics[width=0.85\linewidth]{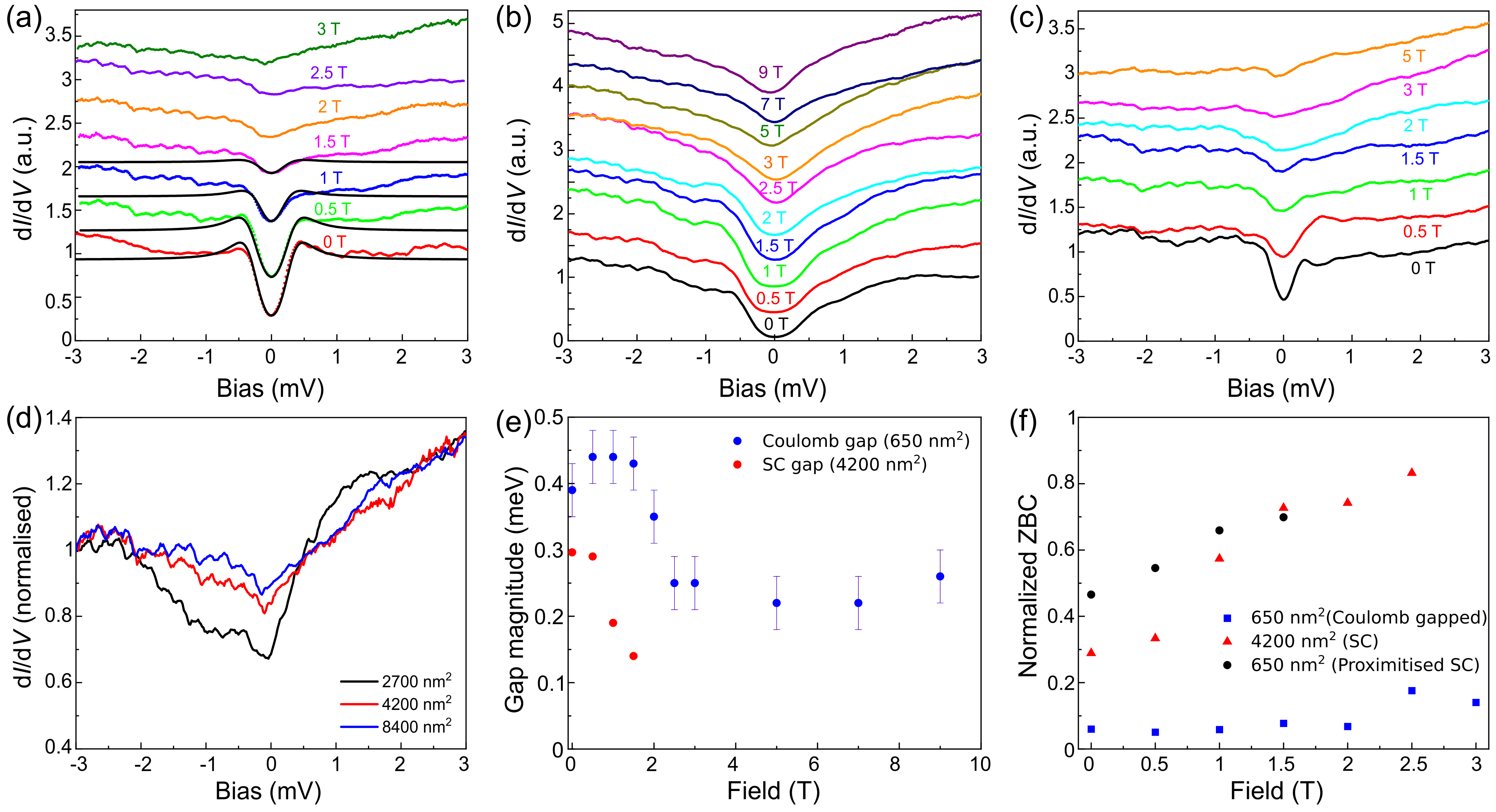}
\caption{Magnetic field dependence of SC gap, Coulomb gap and proximitised SC gap. Magnetic field dependence of d$I$/d$V$ spectra in island of size (a) 4200 nm$^2$ including BCS fits up to 1.5 T, (b) 650 nm$^2$, (c) 650 nm$^2$ having proximity to SC island. (d) Comparison of d$I$/d$V$ taken at 3 T on islands with sizes of 2700, 4200 and 8400 nm$^2$. (e), (f) Magnetic field dependence of (e) gap magnitude and (f) normalized zero bias conductance. Spectra in panels (a),(b),(c) are offset vertically for clarity.}
\label{Field dep}
\end{figure}

In Figure \ref{Field dep}, we characterise and distinguish the different types of observed spectra by their respective magnetic field dependencies. The representative spectra chosen are superconducting (SC) spectra from island of size 4200 nm$^2$ (Fig.~\ref{Field dep}(a)), Coulomb gapped spectra from island of size 650 nm$^2$ (Fig.~\ref{Field dep}(b)), and proximitized SC spectra from island of size 650 nm$^2$ (Fig.~\ref{Field dep}(c)). The Coulomb nature of the island having size 2700 nm$^2$ becomes evident from the d$I$/d$V$ taken at magnetic field of 3 T, which is greater than H$_{c2}$ for the superconducting islands. As shown in Fig.~\ref{Field dep}(d), the superconducting island having size close to the SC-Coulomb phase boundary (4200 nm$^2$) and larger SC island (8400 nm$^2$) have almost indistinguishable d$I$/d$V$ at $H=3$ T whereas in the 2700 nm$^2$ island, significantly prominent gap signature is present easily distinguishable from the larger SC islands. The fitted BCS gap magnitude for island of size 4200 nm$^2$ decreases monotonically to zero with increasing field, whereas Coulomb gap in 650 nm$^2$ island remains finite and almost constant (Fig.~\ref{Field dep}(e)). The normalized zero bias conductance (ZBC) of both SC and proximitized SC spectra monotonically increases to one with increasing field whereas ZBC for Coulomb gap remains almost constant with increasing field (Fig.~\ref{Field dep}(f)).

It is worthwhile to note that even for large SC islands we still have a small residual gap at high magnetic fields which could be attributed to a correlated pseudogap state that appears when the superconductivity is quenched, and can have additional symmetry broken states of charge or spin-ordering. Precise nature of this pseudogap state remains an open question in the field of correlated superconductors. It is noted that the small residual gap in high fields observed in Pb islands was attributed to the Coulomb gap\cite{Yuan2020}.

\newpage
\section{Spatial dependence of point spectra within a single island}
\begin{figure}[h!]
\center
\includegraphics[width=0.98\linewidth]{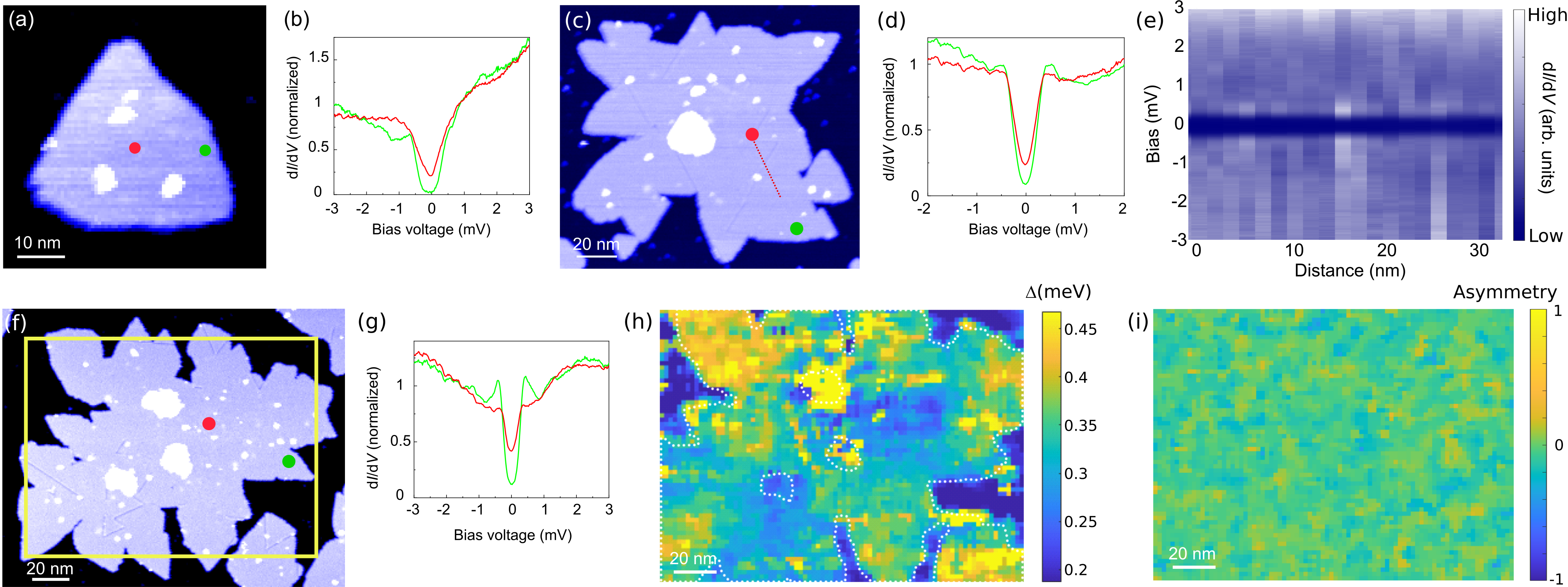}
\caption{Spatial variation of d$I$/d$V$ on island sizes. (a) Island size 650 nm$^2$. (b) Point spectra taken at indicated locations in (a). (c) Island size 10400 nm$^2$. (d) Point spectra taken at indicated locations in (c). (e) Line spectra across the domain wall indicated by the red dotted line in (c). (f) Island size 18400 nm$^2$. (g) Point spectra taken at indicated locations in (f). (h) Superconducting gap map from the Dynes fit of the d$I$/d$V$ map taken at the area indicated by yellow rectangle at (f). The outline of the island's topographic feature is indicated by white dashed line. (i) Spatial asymmetry map.}
\label{fig:Spatial dep}
\end{figure}

In Fig.~\ref{fig:Spatial dep}(a,b), we demonstrate the representative spectra in edge and middle of a Coulomb-gapped island indicating that the magnitude of the Coulomb gap remains constant. Spatial variation of SC spectra is demonstrated in Fig.~\ref{fig:Spatial dep}(d,g) for 2 different sized islands in Fig.~\ref{fig:Spatial dep}(c,f) having areas 10400 nm$^2$ and 18400 nm$^2$, respectively (lateral sizes 102 nm and 136 nm, respectively). It is observed that the edges of the SC islands shows larger SC gap compared to the middle of the islands and this variation is larger for bigger islands. In fig.~\ref{fig:Spatial dep}(e), the linespectra across CDW domain wall in fig.~\ref{fig:Spatial dep}(c) (shown in red dotted line), indicates unchanged SC gap. Fig.~\ref{fig:Spatial dep}(h) shows the spatial variation of the fitted SC gap with Dynes equation for spectra taken in island in Fig.~\ref{fig:Spatial dep}(e). Fig.~\ref{fig:Spatial dep}(i) shows spatial asymmetry: ($\mathrm{d}I/\mathrm{d}V (V=V_\mathrm{cp})-\mathrm{d}I/\mathrm{d}V (V=-V_\mathrm{cp})$, where $V_\mathrm{cp}$ is the bias corresponding to the coherence peak positions).

\section{Superconducting gap map and zero bias conductance map of smallest superconducting island}
\begin{figure}[h!]
\center
\includegraphics[width=0.8\linewidth]{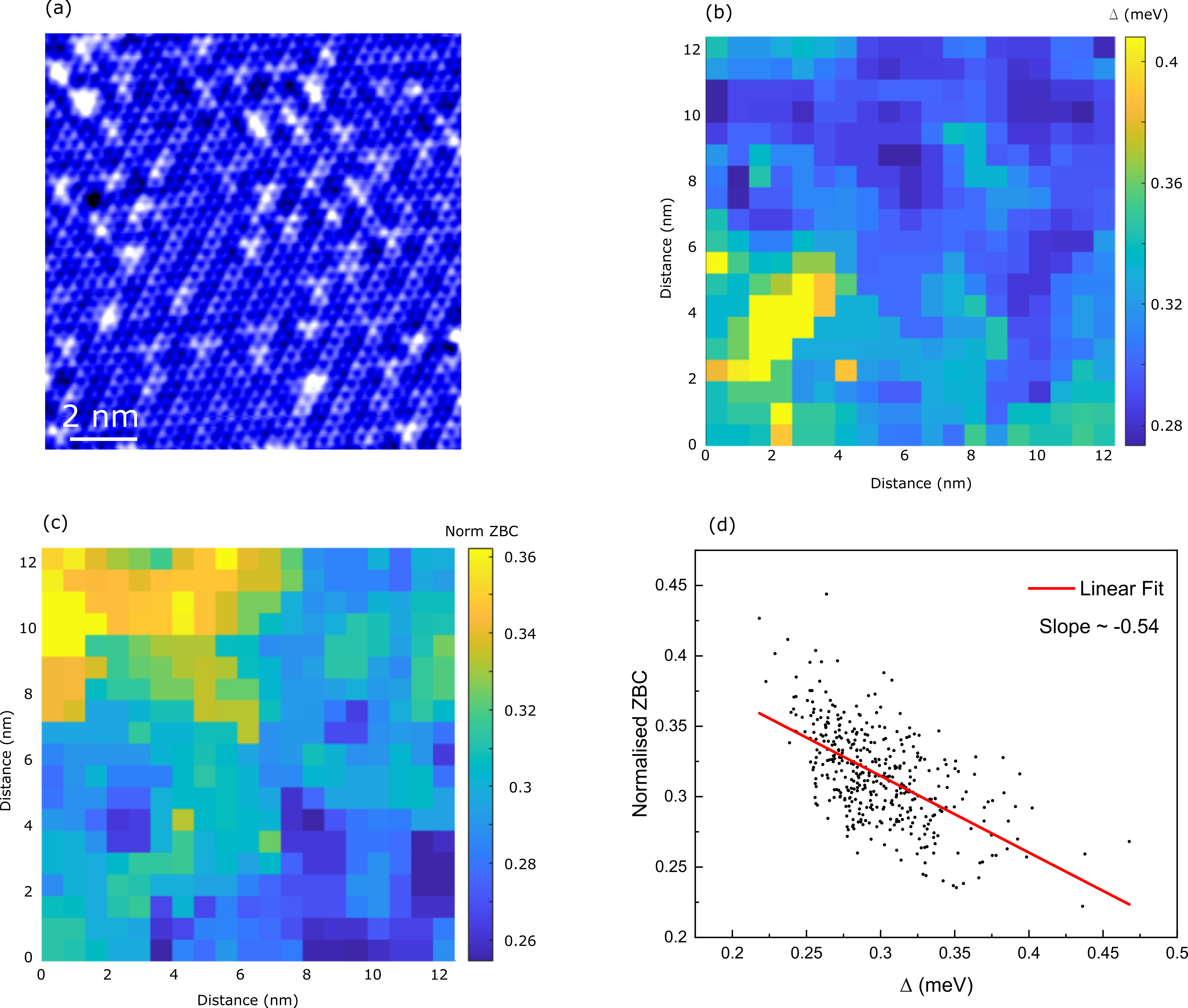}
\caption{Spatial variation of superconducting gap and ZBC in superconducting island of size 4200 nm$^{2}$. (a) Atomic resolution of the area. (b) Spatial variation of fitted SC gap. (c) Spatial variation of ZBC. (d) ZBC vs SC gap along with its linear fit yields a slope of -0.54.}
\label{fig:gapmap}
\end{figure}

For the smallest island on which the superconducting spectra was observed (size $\sim 4200$ nm$^2$), d$I$/d$V$ map was obtained over an area 12.5 nm $\times$ 12.5 nm (Fig.~\ref{fig:gapmap}(a)). The gap obtained from the fitted spectra with Dynes model and the normalised zero bias conductance shows spatial variation as seen in Fig.~\ref{fig:gapmap}(b,c). It is also apparent from the spatial variations that the regions where ZBC is higher, SC gap is lower and vice versa. ZBC vs SC gap plot fit gives a slope of $\sim$ -0.54 indicating strong anticorrelation.

\newpage 

\section{Dependence of the transition on the number of many-body orbitals}

\begin{figure}[h!]
\center
\includegraphics[width=0.6\linewidth]{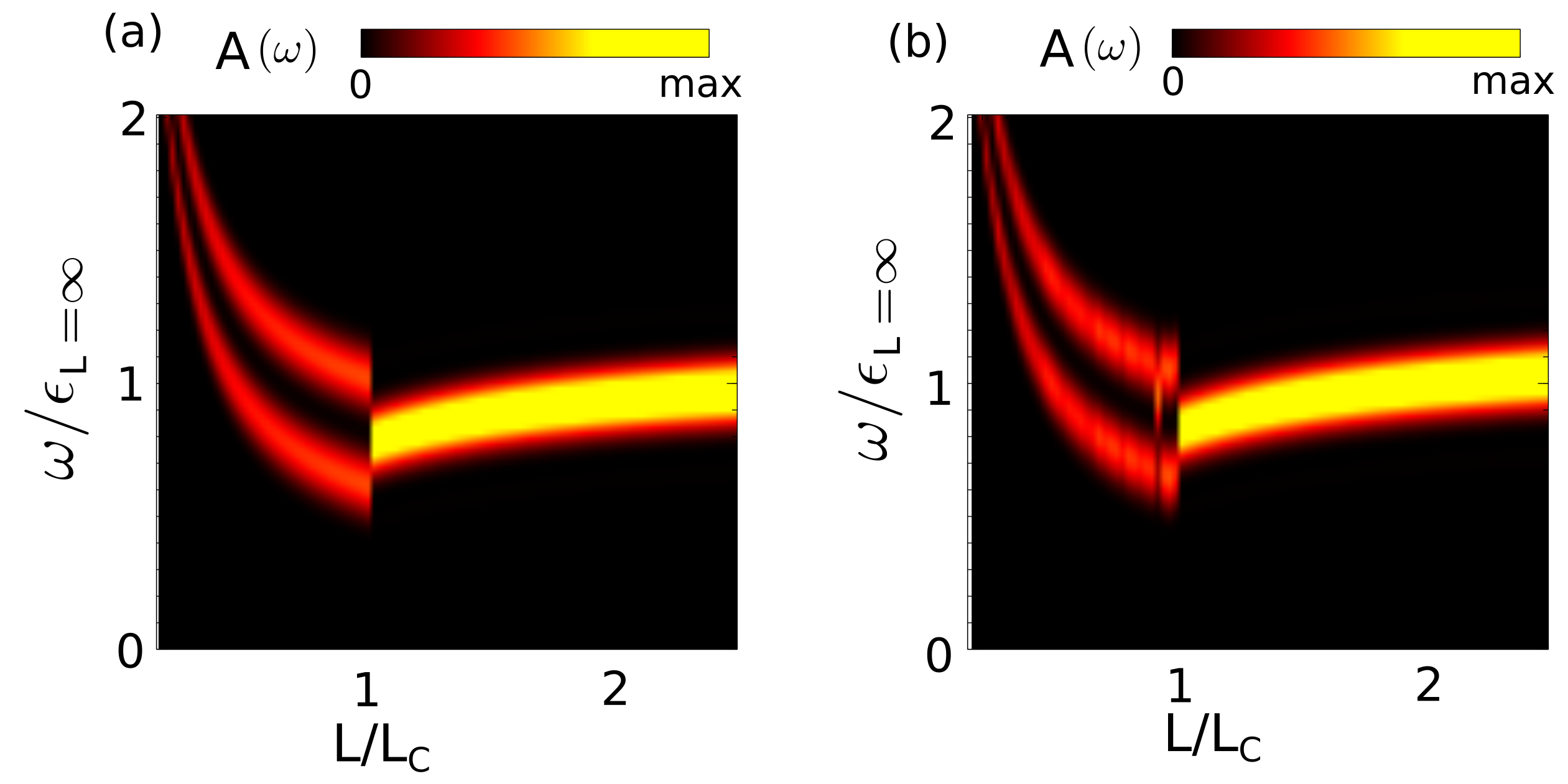}
\caption{Electronic spectral function as a function of the
size of the island, for a different number of many-body
orbitals, $2n=10$ in (a) and $2n=12$ in (b).
It is observed that the transition between
different gaps happens irrespective of the number of orbitals.}
\label{fig:SMorb}
\end{figure}

Here we show that the transition between the correlated gap and the superconducting gap takes
place independently on the number of orbitals considered in the calculation.
In particular, we show in Fig.~\ref{fig:SMorb} the spectral function as a function of the
size of the island, taking a different number of many-body orbitals in the calculations.
It can be clearly seen that both for $2n=10$ (Fig.~\ref{fig:SMorb}(a))
and $2n=12$ (Fig.~\ref{fig:SMorb}(b)) orbitals, a transition between a correlated
gap to a superconducting one emerges, analogous to the 
calculations of Fig.~3(a) in the main manuscript.
In the absence
of $U$ or $V$ there would not be a phase transition between the correlated
and superconducting regimes. Nevertheless,
in the absence of repulsive interactions, 
there could still be a phase transition as function of the system size between 
a single-particle gap coming from quantized energy levels
and a superconducting gap. This transition would have an associated
smooth evolution of the gap, in contrast with the sharp transition we observe in our data.
We finally note that the zero-bias anomaly can be modelled with the model of the dynamical Coulomb blockade\cite{Ingold1992}. While modelling the dynamical Coulomb blockade in this system could be very interesting, it is significantly beyond the scope of this manuscript.

\section{SC-Coulomb phase boundary as a function of the strength of the proximity effect}
\begin{figure}[h!]
\center
\includegraphics[width=0.6\linewidth]{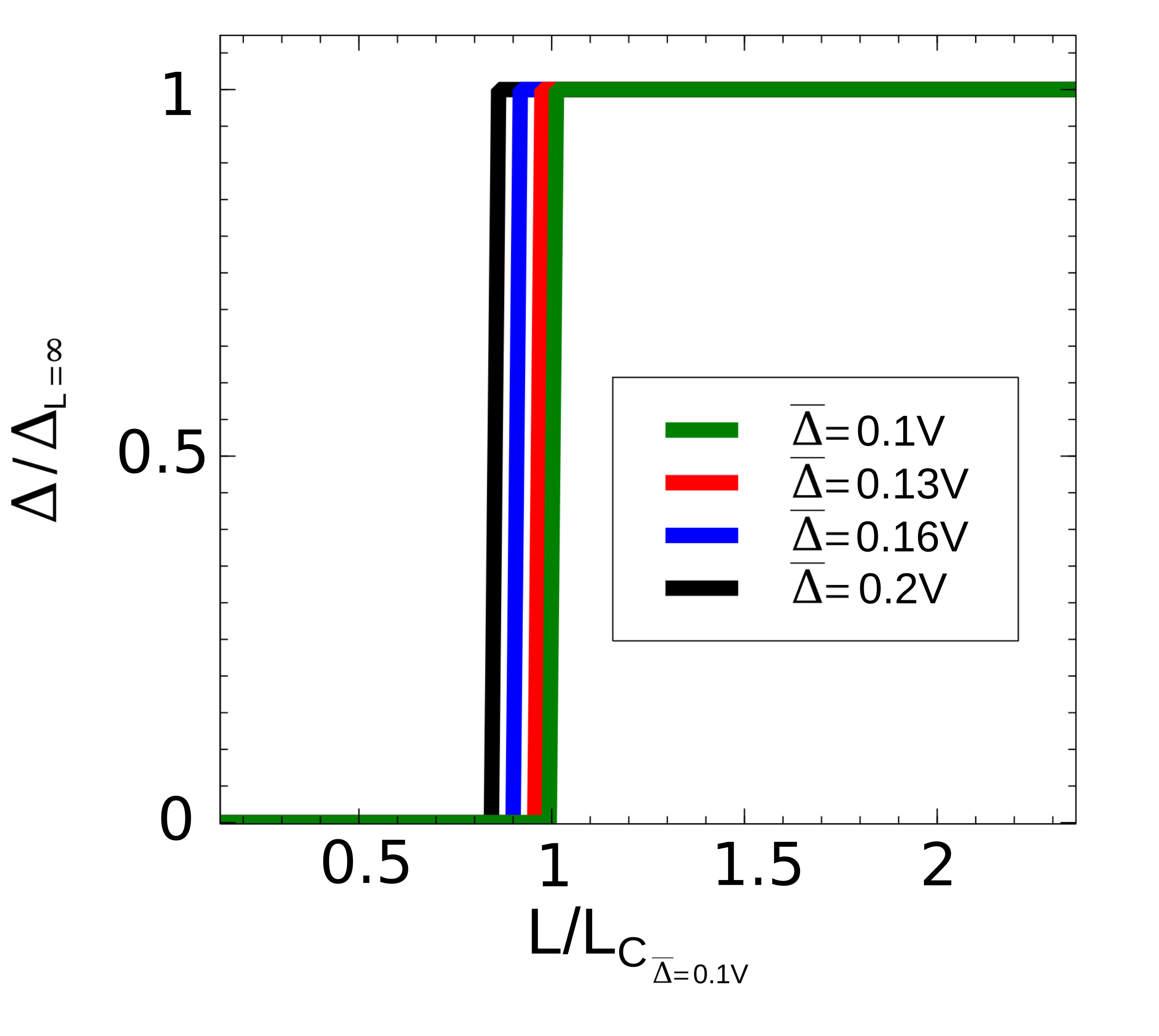}
\caption{Transition from a correlated gap to a superconducting gap as a function of the size of the island for increasing values of the superconducting proximity effect.}
\label{fig:Phase}
\end{figure}

Here we address the dependence of the transition between the correlated and superconducting state driven by proximity. First, it is worth noting that in the absence of superconducting proximity, the NbSe$_2$ would not show the presence of superconductivity due to its finite nature. However, for intermediate islands, the existence of a small proximity drives the system to the superconducting state. The critical
length at which such transition takes place depends on the strength of the proximity effect as shown in Fig.~\ref{fig:Phase}. The transition remains sharp for finite proximity effects, illustrating that a sharp transition with system size is expected irrespective of the exact value of the superconducting proximity effect.

\clearpage

\section{Coulomb-enhanced interactions including charge-density wave and spin-orbit coupling}
NbSe$_2$ is well known for having a complex electronic structure combining Ising spin-orbit coupling
and charge density wave. In that regard, it is worth considering whether those effects would have
a non-trivial interplay with the confinement enhanced interaction. In this section, we address this issue,
showing that Ising spin-orbit coupling nor charge density wave qualitatively change the picture
presented in the main manuscript.

In the following we will consider
an atomistic tight binding model for the Wannier states
of the NbSe$_2$ band, one per Nb atom, sitting
in a triangular lattice \cite{Smith1985}. The total Hamiltonian takes the form
\begin{equation}
H = 
H_{\text{kin}} + 
H_{\text{CDW}} + 
H_{\text{SOC}}
\end{equation}
where $H_{\text{kin}}$ is the spin-independent hopping term
\begin{equation}
    H_{\text{kin}} = 
    \sum_{i,j,s } t_{ij} c^\dagger_{i,s} c_{i,s}
\end{equation}
$H_{\text{CDW}}$ is the charge density-wave order
\begin{equation}
    H_{\text{CDW}} = 
    \sum_{i,s } \epsilon_{\text{CDW},i} c^\dagger_{i,s} c_{i,s}
\end{equation}
and $H_{\text{SOC}}$ is the intrinsic Ising spin-orbit coupling
\begin{equation}
    H_{\text{SOC}} = 
    i\lambda_{\text{SOC}}\sum_{\langle ij \rangle,s,s' } \gamma_{ij} \sigma_z^{s,s'} c^\dagger_{i,s} c_{i,s'}
\end{equation}

\begin{figure}[t!]
\center
\includegraphics[width=0.6\linewidth]{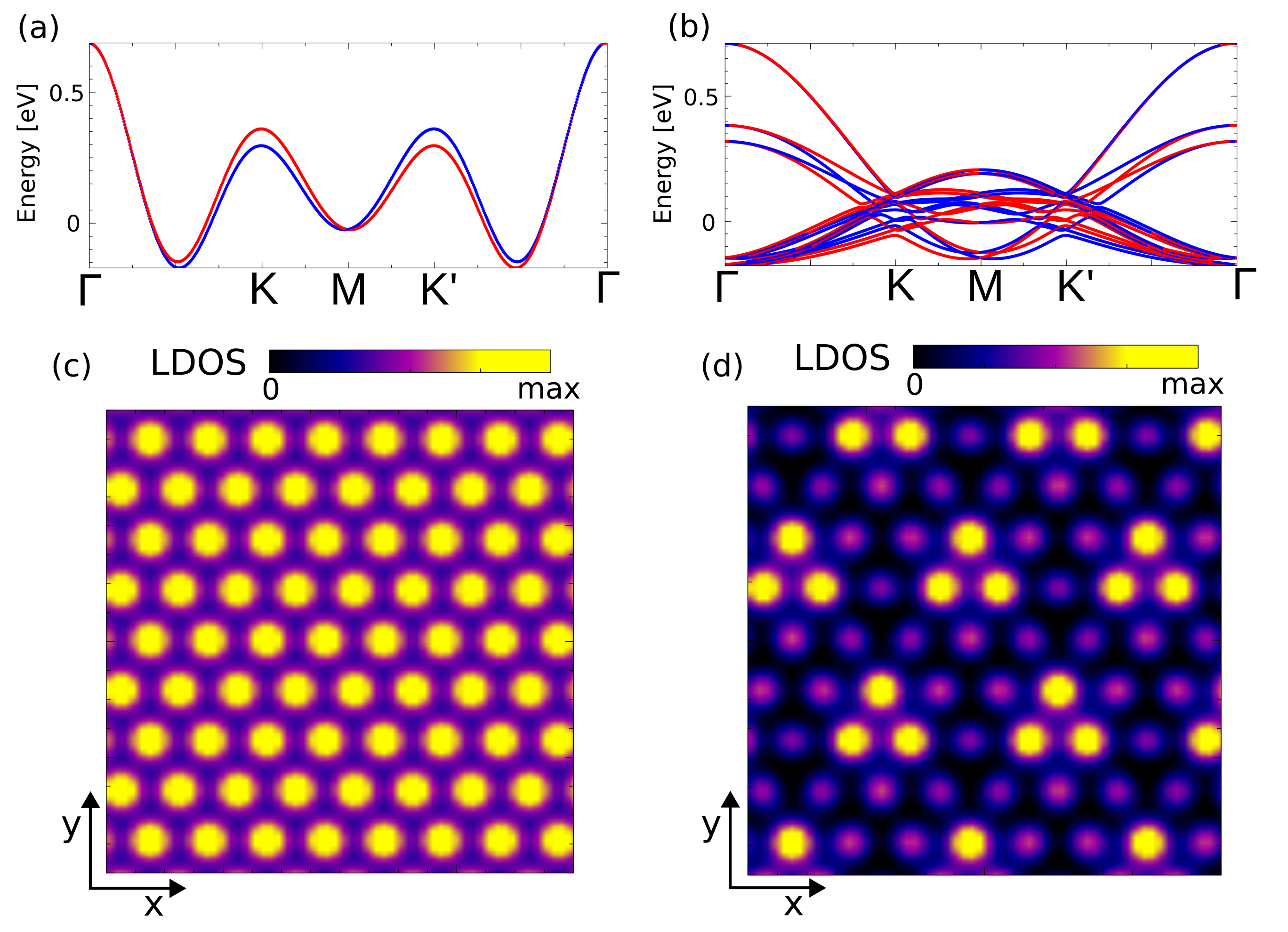}
\caption{Electronic structure of NbSe$_2$.
Band structure for the minimal unit cell in the absence of CDW (a), and LDOS (c).
Panel (b) shows the band structure in the presence of CDW for a 3x3 supercell,
and panel (d) shows the LDOS reflecting the spatial charge modulation of the
CDW.}
\label{fig:SM:TB}
\end{figure}
In the previous terms, $t_{ij}$ is the hopping term that incorporates up to
4th-neighbor hopping, $\epsilon_{\text{CDW},i}$ are the modulated onsite energies associated
to the CDW order,
$\nu_{ij}=\pm 1$ alternate signs between the different bonds
leading to a $C_3$ symmetric hopping \cite{PhysRevLett.95.226801},
and $\sigma_z$ is the spin Pauli matrix. The band structure and local density
of states at Fermi energy are shown in Fig.~\ref{fig:SM:TB}. It is observed that
the bands feature the momentum dependent spin-splitting associated to the
Ising spin-orbit coupling, and that in the presence of the CDW perturbation the
local density of states (LDOS) reproduces the experimentally observed features. As a result,
the previous model captures all the microscopic features of NbSe$_2$.

With the previous Hamiltonian, we now consider the effect of interactions on a finite island.
We include interactions in the form of long-range Coulomb interaction in the atomistic model as
\begin{equation}
    H_{\text{Coulomb}} = \sum_{i\ne j,s,s'}
    \frac{V_0}{|\mathbf{r_i} - \mathbf{r_j} |}c^\dagger_{i,s} c_{i,s} c^\dagger_{j,s'} c_{j,s'}
    \label{eq:coulomb}
\end{equation}
where $\mathbf{r_i}$ is the location of Nb atom $i$,
and $V_0$ is the Coulomb prefactor in atomic units. As this term turns the system into a full-fledge many-body problem,
we will consider the impact of interactions only in the lowest energy states.
We project this Coulomb interaction to the
states closest to the Fermi surface $\Psi_\alpha$, giving rise to an interaction of the form

\begin{equation}
     \mathcal{H}_{\text{Coulomb}} = \sum_{ijkl} \mathcal{V}_{ijkl} \Psi^\dagger_{i} \Psi_{j} \Psi^\dagger_{k} \Psi_{l}
\end{equation}

\begin{figure}[b!]
\center
\includegraphics[width=0.6\linewidth]{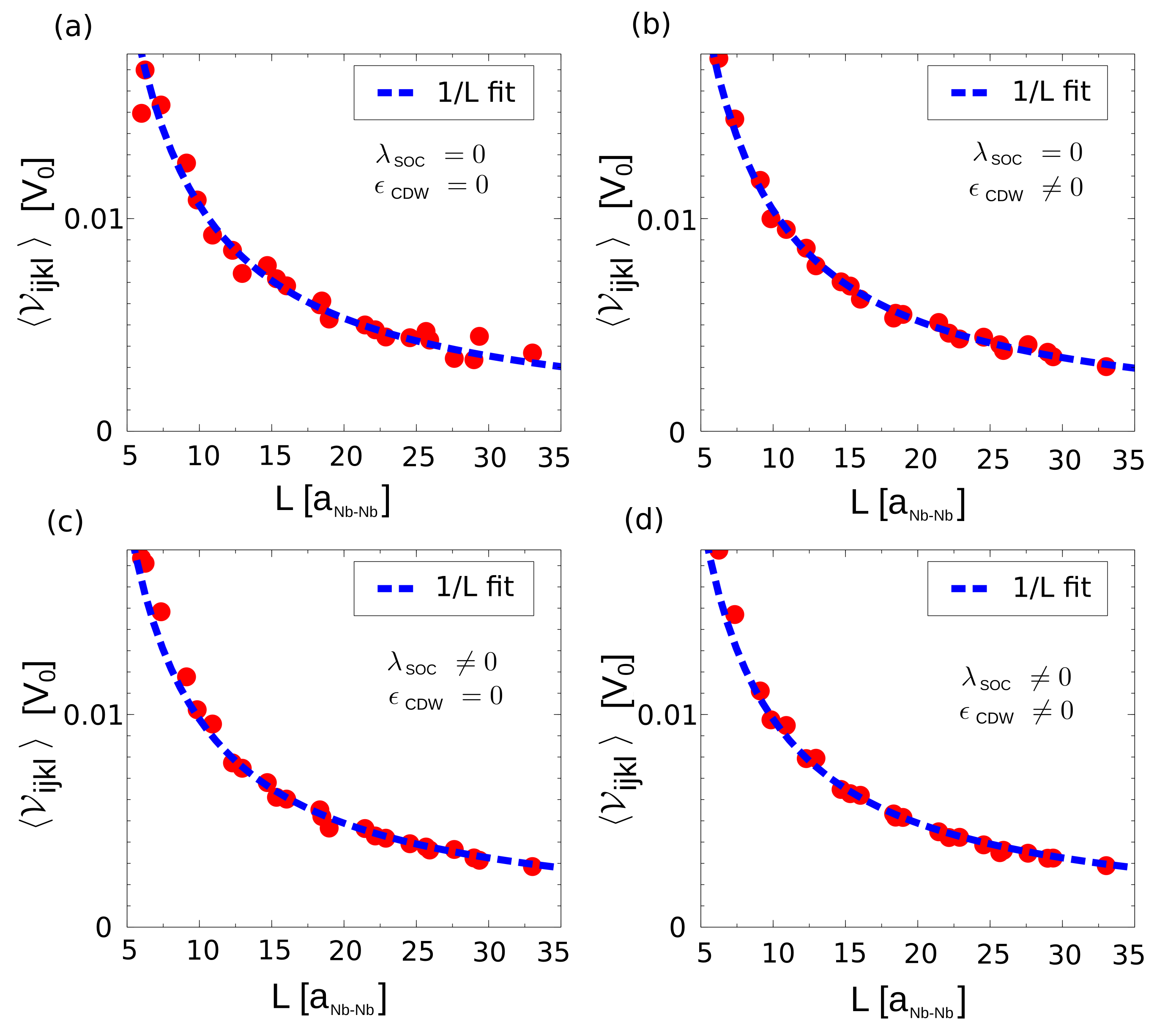}
\caption{Confinement controlled Coulomb interactions.
Average value of the projected interaction $\mathcal{V}_{ijkl}$
on the Fermi surface states $\Psi_\alpha$.
The projected interaction is computed for model zero SOC and zero CDW (a),
zero SOC and finite CDW (b), zero CDW and finite SOC (c), and finite CDW and finite SOC
(d) It is observed that the average repulsive interaction shows a $1/L$
dependence, with $L$ the size of the island, in agreement the model
in the main manuscript.
}
\label{fig:SM:Cou}
\end{figure}
where $\mathcal{V}_{ijkl}$ are obtained by projecting Eq. \ref{eq:coulomb} into the low energy states $\Psi_\alpha$.
We now take finite-size islands with different shapes and different number of atoms, and compute
the effective interaction $\mathcal{V}_{ijkl}$. We perform this procedure on islands whose atomistic Hamiltonian
has zero and non-zero spin-orbit coupling and zero and non-zero charge density wave, the
results are shown in Fig. \ref{fig:SM:Cou}. As the effective interaction $\mathcal{V}_{ijkl}$ is a four dimensional
tensor, generically complex-valued, 
we will characterize the strength of the repulsive interactions by the average of its absolute value.
As shown in Fig.~\ref{fig:SM:Cou}, we observe a robust $1/L$ behavior of the projected interaction, independently
on the presence or absence of Ising spin-orbit coupling and charge density wave. These results demonstrate that
not Ising spin-orbit coupling nor the charge density wave create an impact on the projected interactions and that
the repulsive interactions are purely dominated by the island confinement effect.

We now comment on the possibility of extracting the exact parameters for our theoretical model
from the experimental data.
Extracting $U_0$ and $c_0$ and the
full dependence of $L_\mathrm{C}$ on $U_0$, $c_0$, $V$, $\mu$, and $\bar \Delta$
would require a very large number of islands so
that a detailed error analysis can be done when performing the fitting.
Otherwise, small fluctuations would give rise to an inaccurate estimate of $U_0$ and $c_0$.
Since at the current stage, our experiments do not allow for a precise estimate of those
values, our discussion is focused on highlighting the qualitative behavior of the system
rather than providing a specific quantitative extraction of the parameters.
From the materials point of view, it is
worth noting that $U$ and $V$ are effective interactions projected onto the low energy
states, that have a highly non-trivial dependence on the different material parameters of NbSe$_2$. Furthermore,
details such as surface effects and substrate effects could impact these parameters.

Finally, we comment on interaction screening effects in NbSe$_2$. From the theoretical point of view, providing an accurate estimate of the screening would require performing RPA (random phase approximation) density functional theory (DFT) calculations of the dielectric screening, which would account for both intraband and interband screening in the system. We note that such estimate cannot be reliably performed with the low energy tight binding model we are considering, as interband contributions would be completely neglected in that scenario. 

\clearpage

\section{Estimation of coherence length}

\begin{figure}[h!]
\center
\includegraphics[width=0.95\linewidth]{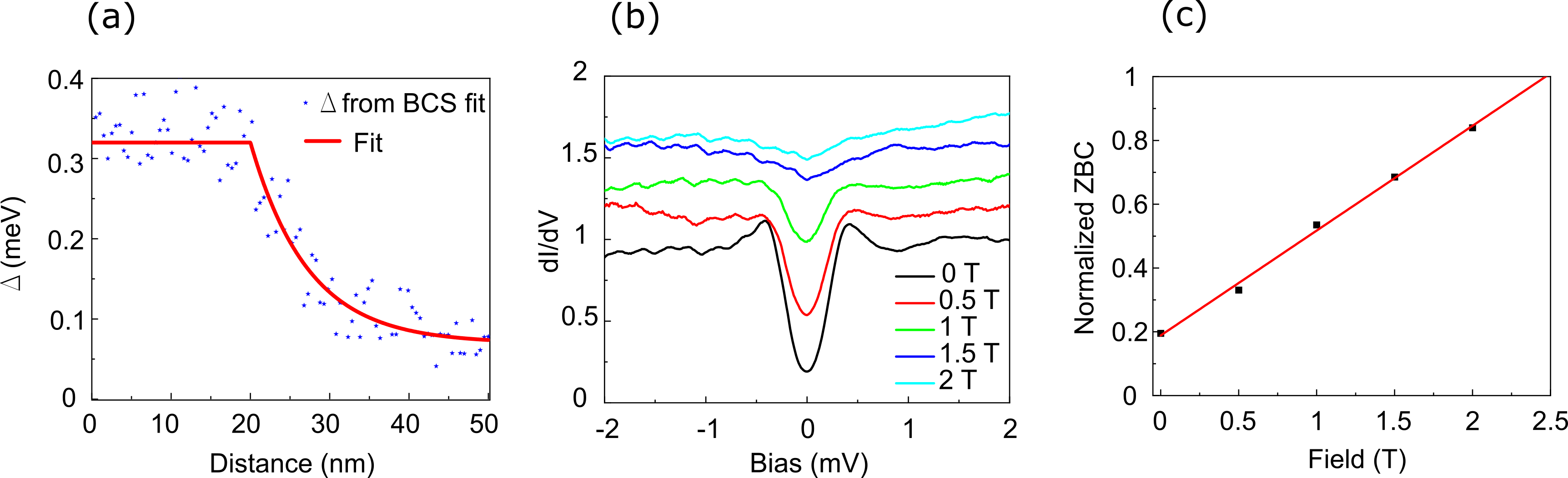}
\caption{Estimation of Coherence length. (a) Exponential fit of proximitised SC gap. (b) Magnetic field dependence of SC spectra. (c) Linear interpolation of normalized ZBC.}
\label{Coherence}
\end{figure}

To estimate the superconducting coherence length, we first fitted the  proximitized spectra of Fig. 4 (c) with Dynes' model. The extracted gap values as function of distance was then plotted with an exponential decay to extract the Coherence length in Fig. ~\ref{Coherence}(a). The fitting formula was $\Delta(x)=\Delta_{1}-\Delta_{2}(1-e^{\frac{-(x-x_{0})}{\xi}})$, where $\Delta(x), \Delta_{1}, (\Delta_{1}-\Delta_{2}), x_{0}, \xi$ are fitted gap, SC gap inside the NbSe$_{2}$ island, residual SC gap in HOPG (due to proximity of nearby islands), spatial location of the boundary between NbSe$_{2}$ and HOPG and the Coherence length respectively. This fit gives us a Coherence length $\xi$ $\approx$ 7.3  nm.

Alternatively, the coherence can be determined from the field dependence of the SC island shown in Fig. ~\ref{Coherence}(b). The SC island size here is 8400 nm$^{2}$. From the linear interpolation of the normalized zero bias conductance (ZBC), we can estimate the upper critical field ($H_{C2}$), which comes out to be $\approx$ 2.47 T. So, the estimated coherence length will be $ \xi = \sqrt {\frac{1}{2\pi}\frac{\phi_{0}}{H_{c2}}}$ = 11.5 nm.  

\newpage
\section{Spatial dependence of islands in proximity}
\begin{figure}[h!]
\center
\includegraphics[width=0.90\linewidth]{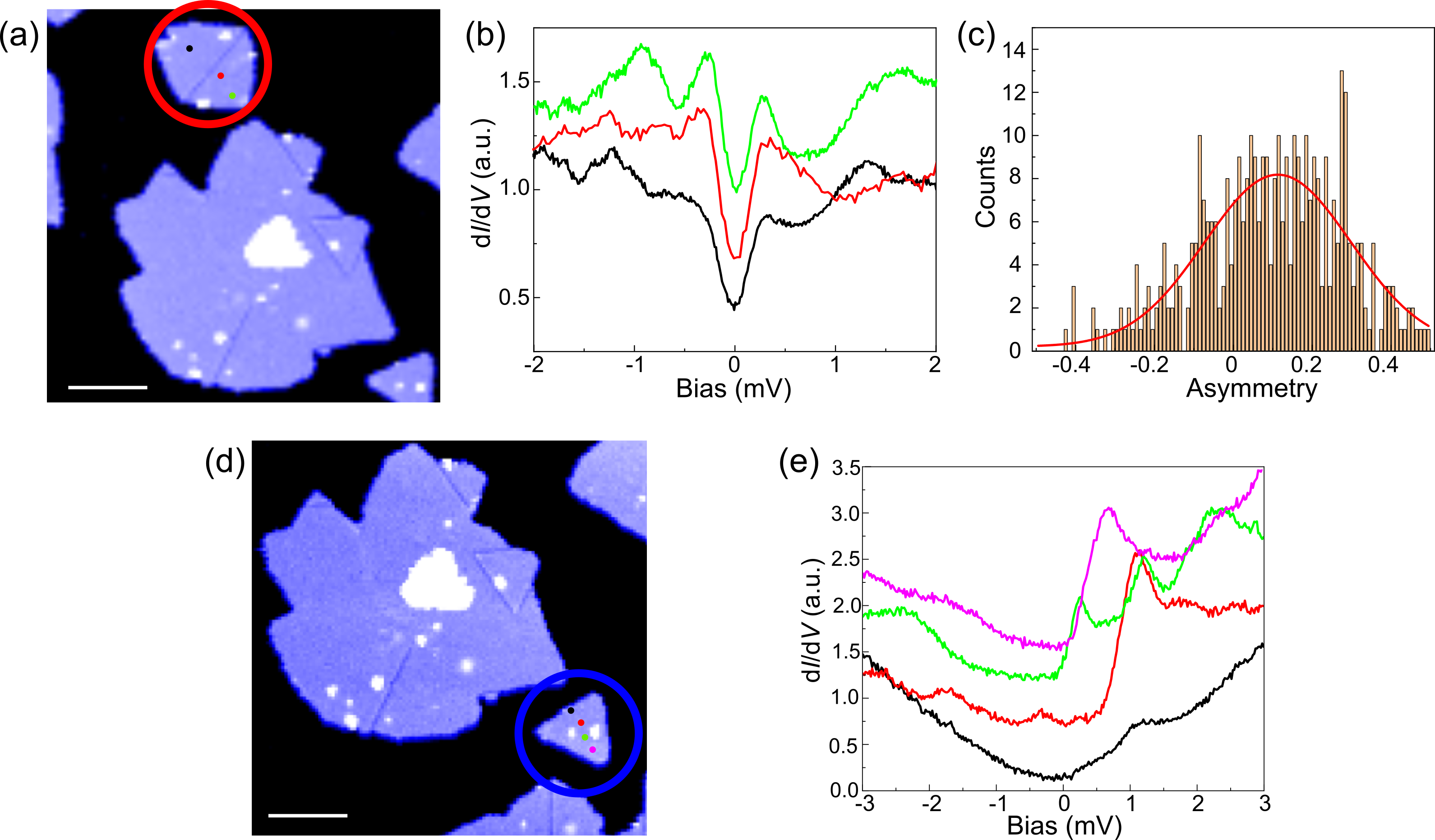}
\caption{Spatial variation of proximitised spectra. (a) Proximitised superconducting island. Scale bar, 20 nm. (b) Point spectra at the indicated locations. (c) Statistical distribution of the asymmetry in the individual spectra and its Gaussian fit. (d) Proximitised non-superconducting island. Scale bar, 20 nm. (e) Point spectra at the indicated locations.}
\label{Proximitised}
\end{figure}

The proximity induced SC gap in the island in Fig.~\ref{Proximitised}(a) varies spatially as illustrated in Fig.~\ref{Proximitised}(b). We observe the presence of SC order by the dip in the conductance at zero bias and the presence of coherence peaks in the different locations of the proximitised island. It indicates that the SC order has been established in the entire island. The SC spectra is asymmetric in conductance values at the coherence peak locations. There is however a variation in the asymmetry observed in the individual spectra at different locations. The histogram of the asymmetry defined by the (normalised conductance at -ve coherence peak location)-(normalised conductance at +ve coherence peak location) shows a distribution asymmetric about zero (mean value of 0.11 from Gaussian fit). The proximitised non-SC island in Fig.~\ref{Proximitised}(d) have variations in the local spectra as seen from Fig.~\ref{Proximitised}(e). Here, the strong electron-hole asymmetries are typical of a non-superconducting origin of the gap \cite{Cai2016}, and support the strongly correlated nature of the small islands. 

\bibliography{biblio}{}